\documentclass[12pt]{article}
\usepackage[cp1251]{inputenc}
\usepackage{amsfonts}
\usepackage[english]{babel}
\usepackage{eufrak}
\usepackage{cite}
\usepackage{amssymb}

\begin{document}

\newcommand{\gguide}{{\it Preparing graphics for IOP journals}}
\title{An Illustration of Generalized Thermodynamics by Several Physical Examples}

\author{ D.V. Prokhorenko \footnote{prokhordv@yandex.ru}} \maketitle

\begin{abstract}
It has been shown recently that Bose Gase with weak pair (enough
well) interaction is non ergodic system. But Bose Gase with weak
pair interaction is so general system that it is evident that the
majority of statistical mechanics systems are non ergodic too. It is
also has been shown that it is possible to generalize the scheme of
standard statistical mechanics and thermodynamics to take into
account non ergodicity. This generalization is called a generalized
thermodynamics. In some points this generalized thermodynamics
coincide with standard equilibrium thermodynamics but some new
specific results take place. It has been shown that this new
generalized thermodynamics can be used to explain some physiological
phenomena which take place in the living cell when the cell is
exciting and dying.

In the present paper we try to illustrate some basic points of this
generalized thermodynamics on some physical examples.
\newpage

\end{abstract}
\sloppy
\section{Introduction} The purpose of this paper to give some
examples on so called generalized thermodynamics recently developed
in [1,2].

In [1] it has been shown that even most realistic systems of
statistical mechanics (for example usual Bose Gase with weak (enough
well) pair interaction) are non ergodic systems. These means that
there exist non trivial first integrals of the system commuting with
momenta and number particle operatots. It has been recently shown
[2] that it is possible to generalize the scheme of standard
statistical mechanics and thermodynamics to take into account the
existence of such non trivial first integrals. These generalization
we call the generalized thermodynamics. It has been also shown that
this generalized thermodynamics is very useful to explanation of
some physiological phenomena which take place in the living cell
when the cell is excited dying.

Let us describe the basic elements of the scheme of this generalized
thermodynamics. The non ergodic theorem states that for a wide class
of realistic systems of statistical mechanics there exists non
trivial commuting (in involution) first integrals \(K_1,...,K_N\),
\(N=1,2,...\) commuting with momenta and number particle operators.
In purpose of simplicity we will talk about Hamiltonian instead of
Hamiltonian, momenta and number particle operators. The starting
point of generalized thermodynamics is a following expression for
distribution function (density matrix in quantum case):
\begin{eqnarray}
\rho(x)=\rm const \mit\;\delta(H(x)-E)\prod
\limits_{i=1}^{N}\delta(K_i(x)-K'_i), \label{I}
\end{eqnarray}
where \(x\) is a point of phase space of the system and
\(E,\;K'_1,...,K'_N\) are the observable values of energy and
integrals \(K_1,...,K_N\). The entropy corresponding to this
distribution is defined as a logarithm of statistical weight. The
statistical weigh, by definition, is a number of microscopic
configuration of the system, corresponding to a given macroscopic
state i.e.:
\begin{eqnarray}
W(E,K'_1,...,K'_N)=\int d\Gamma_x \delta(H(x)-E)\prod
\limits_{i=1}^{N}\delta(K_i(x)-K'_i)\label{IIa}
\end{eqnarray}
and
\begin{eqnarray}
S(E,K'_1,...,K'_N)=\ln W(E,K'_1,...,K'_N),\label{IIb}
\end{eqnarray}
where \(d\Gamma_x\) is an element of phase volume. The expression
(\ref{I}), (\ref{IIa}) and (\ref{IIb})  are the generalization of
standard microcanonical Gibbs distribution and its entropy to the
case when there exists non trivial commuting first integrals
\(K_1,...,K_N\) of the system.

Let us explain why we require pairwise commutativity of integrals
\(K_1,...,K_N\) or why we require that \(\forall i,j=1,...,N\)
\((K_i,K_j)=0\) where \((\cdot,\cdot)\) denotes the commutator in
quantum case or the Poison bracket in classical case. We want to
describe the state with definite values \(K'_1,...,K'_N\) of first
integrals \(K_1,...,K_N\). So the integrals \(K_i,\;i=1,...,N\) must
be contemporaneously measurable. But in quantum mechanics this means
that the integrals \(K_1,...,K_N\) must be pairwise commutative. The
requirement that the integrals \(K_1,...,K_N\) must be in involution
(in classical case) is clear now from the remark that the Poison
bracket is a classical analog of commutator.

Now let answer the question when generalized thermodynamics gives
results which differs from results following from standard
thermodynamics. I.e. let us answer the question when using of
distribution (\ref{I}) leads to results which differ from result
obtained by using standard microcanonical Gibbs distribution.

For simplicity consider the case \(N=1\). The case of an arbitrary
\(N\) can be considered by analogy to this case. For a fixed \(E\)
\(S(E,K')\) is a function of \(K'\). There exist two typical cases
of behavior of this function.

1) \(S(E,K')\) (for fixed \(E\)) has a maximum in isolated point
\(K'=K''\).

2) \(S(E,K')\) achieve a maximum at whole interval \(K' \in [a,b]\)
of nonzero length.

There arise a question, which observable values \(K'\) of integral
\(K\) could be realized in nature. It is essentially to suppose that
it could be realized such and only such values \(K'\) of \(K\) which
corresponds to the maximum of entropy (for fixed energy). This is
one of main point of generalized thermodynamics. It has been shown
[2] that in case 2) generalized thermodynamics gives no more than
standard equilibrium thermodynamics (\(K'\)) is uniquely defined).
But new interesting physics arise in case 2). Note that the
situation when thermodynamical functions have a plateau as functions
of their arguments is a typical situation in the theory of phase
transitions.

Some general results of standard equilibrium thermodynamics take
place in generalized thermodynamics [2]. For example a well-known
relation
\begin{eqnarray}
dE=TdS-PdV
\end{eqnarray}
takes place in new situation. Here \(E\) is an energy of the system,
\(T\) is a temperature of the system, \(V\) is a volume of the
system, \(P\) is a pressure. But some new specific results take
place for generalized thermodynamics. These new results have been
used [2,3] for explanation of several physiological phenomena which
takes place in biological cell when the cell is dying or moving from
resting state into activated state.

In the present paper we give some new physical examples to
illustrate some basic points of this generalized thermodynamics. The
paper is composed as follows. In section 3 we describe some new
derivation of the Gibbs distribution from the property of
asymptotical factorization of correlations based on ideas of
nonequilibrium renormalization theory. In section 3  we reformulate
the Bardeen --- Cooper --- Schrieffer model on the language of the
generalized thermodynamics.In section 5 we describe some derivation
of the Boltzmann kinetic equation based on some ideas of N.N.
Bogoliubov concerned with thermalization in oscillator interacting
with thermostat.  In section 6 we describe how to obtain non-trivial
first integrals for the equations on order parameter of the system
in many-phase domain in the theory of second order phase
transitions. Section 6 is a conclusion.

\section{Derivation of Gibbs distribution from the property of asymptotic space factorization of correlations.}
In the present section we prove that all stationary
translation-invariant states of Bose Gase with weak (enough good)
pair interaction satisfying to the property of (enough fast)
asymptotic space factorization of correlations are the Gibbs states
(described by the Gibbs distribution).

But at first let us recall the main result of the paper [1] and
demonstrate, how the non ergodic property follows from this result.

The main result of [1] can be formulated as follows:

\textbf{Theorem 1}. For Bose Gase with weak pair interaction with
kernels from Schwartz space in the sense of formal power series on
coupling constant there exists non-Gibbs functional \(\langle \cdot
\rangle\), commuting with the number of particle operator such that
the correlators
\begin{eqnarray}
\langle\Psi^\pm(t,{x}_1)...\Psi^\pm(t,{x}_n)\rangle \nonumber
\end{eqnarray}
are translation invariant, do not depend on \(t\) and satisfy the
weak cluster property. Here \(\Psi^\pm\) are the secondary quantized
wave function and its complex conjugated \(\Psi^{-}(x):=\Psi(x)\)
and the weak cluster property means the following
\begin{eqnarray}
\lim_{|{a}|\rightarrow\infty}\int \limits_{{R}^{3n}}
\langle\Psi^\pm(t,{x}_1+\delta_1 e_1{a})
...\Psi^\pm(t,{x}_n+\delta_n e_1 {a} )\rangle f({x}_1,...,{x}_n)d^3x_1...d^3x_n\nonumber\\
=\int \limits_{{R}^{3n}}
\langle\Psi^\pm(t,{x}_{i_1})...\Psi^\pm(t,{x}_{i_k})\rangle
\langle\Psi^\pm(t,{x}_{i_k})...\Psi^\pm(t,{x}_{i_n})\rangle
\nonumber \times f({x}_1,...,{x}_n)d^3x_1...d^3x_n,
\end{eqnarray}
there \(\delta_i\in\{1,0\},\;i=1,2...n\) and
\begin{eqnarray}
i_1<i_2<...<i_k,\nonumber\\
i_{k+1}<i_{k+2}<...<i_n,\nonumber\\
\{i_1,i_2,...,i_k\}=\{i=1,2...n|\delta_i=0\}\neq\emptyset,\nonumber\\
\{i_{k+1},i_{k+2},...,i_n\}=\{i=1,2...n|\delta_i=1\}\neq\emptyset.\nonumber
\end{eqnarray}
\(f({x}_1,...,{x}_n)\) is a test function (i.e. the function from
the Schwartz space), \(e_1\) is a unit vector parallel to the
\(x\)-axis. About secondary quantization see for example [4, 5].

Let us prove that the existence of such functionals implies
non-ergodic property of the system. The more accurate proof of this
fact see in section 10 of [1].  Suppose that our system is ergodic,
i.e. there are no first integrals of the system except energy. Then,
the density matrix \(\rho\) of the system corresponding to the
functional \(\langle\cdot\rangle\) is a function of energy. We can
represent this density matrix \(\rho\) as follows:
\begin{eqnarray}
\rho=\sum c_\alpha \delta(H-E_\alpha),\nonumber
\end{eqnarray}
where \(H\) is a Hamiltonian of the system and the sum can be
continuous (integral). Let 1 be some enough large but finite
subsystem of our system. Let 2 be a subsystem obtained from 1 by
translation on the vector \(\vec{l}\) of sufficiently large length
parallel to the \(x\)-axis. Let 12 be a union of the subsystems 1
and 2. Let \(\rho_1\), \(\rho_2\) and \(\rho_{12}\) be the density
matrix of the subsystems 1, 2 and 12 respectively. By the same
method as the method used for the derivation of the Gibbs
distribution we find:
\begin{eqnarray}
\rho_{12}=\sum c_\alpha
d_\alpha\frac{e^{-\frac{H_1}{T_\alpha}}}{Z_\alpha}\otimes\frac{e^{-\frac{H_2}{T_\alpha}}}{Z_\alpha},\;d_\alpha>0\,\forall\alpha
\nonumber
\end{eqnarray}
 in the obvious notation. Here \(H_1\) and \(H_2\) be the Hamiltonians of subsystems 1 and 2 respectively.
 But the weak cluster property implies that
 \begin{eqnarray}
 \rho_{12}=\rho_1\otimes\rho_2.\nonumber
 \end{eqnarray}
 Therefore all the coefficients \(c_\alpha\) are equal to zero except one. We find that
 \begin{eqnarray}
 \rho=c\delta(H-E_0)\nonumber
 \end{eqnarray}
 for some constants \(c\) and \(E_0\). So each finite subsystem of our system can be described by Gibbs
 formula and we obtain a contradiction.

 But the state (not necessary positive defined) mentioned in theorem 1
 satisfies only to the weak cluster property i.e. to the property of
 asymptotic space factorization of correlations only "in one direction".

 It follows from the proof of theorem 1 from [1] that we can achieve,
 that \(\langle\cdot\rangle\) will satisfies to the property of
 asymptotic factorization of correlations "in two directions". But
 there arise principled difficulties if one try to prove that the
 state \(\langle\cdot\rangle\) satisfy to cluster property in "all
 (three)
 directions".

 It has been mentioned above that all stationary
 translation-invariant states satisfying to (enough fast) cluster
 property are the Gibbs states. Let us now recall the standard
 derivation of this fact. Then we will point out some problems
 connected with this derivation and present our new proof of this
 fact based on results of our renormalization theory of
 nonequilibrium (Keldysh) diagram technique.

 Let us at last describe the standard derivation of Gibbs
 distribution (from cluster property) at classical level [6]. This
 derivation is based on the fact that there no exist additive first
 integrals of the system (linear) independent of Hamiltonian \(H\),
 momenta \(\vec{P}\) and angular momenta \(\vec{M}\). Let \(\rho\)
 be a stationary translaion-invariant distribution function of the
 system satisfying to the cluster property of the system. If
 the system is divided into two subsystems 1 and 2 cluster property
 implies that
 \begin{eqnarray}
 \rho=\rho_1\rho_2,
 \end{eqnarray}
 where \(\rho_1\) and \(\rho_2\) are distribution functions for
 subsystems 1 and 2 respectively. In other word the distributions
 for subsystems 1 and 2 are independent. Therefore \(\ln \rho\) is
 an additive integral of motion and can be represented as linear
 function of \(H\), \(\vec{P}\), \(\vec{M}\). In other words
 \begin{eqnarray}
 \ln \rho=\alpha H+\vec{\beta}\vec{P}+\vec{\gamma}\vec{M}
\end{eqnarray}
for some real number \(\alpha\) and real vectors \(\vec{\beta}\) and
\(\vec{\gamma}\). Therefore
\begin{eqnarray} \rho=e^{\alpha
H+\vec{\beta}\vec{P}+\vec{\gamma}\vec{M}}.
\end{eqnarray}
But we assumed that \(\rho\) is translation-invariant. Therefore
\(\vec{\gamma}=0\) and
 \begin{eqnarray}
\rho=e^{\alpha H+\vec{\beta}\vec{P}} \label{Gibbs}
\end{eqnarray}
The distribution (\ref{Gibbs}) is a standard Gibbs distribution. But
there are two problems with this derivation. The first one, this
derivation is performed only at classical level. The second one, we
could not find anywhere the proof of the fact that Hamiltonian \(H\)
momenta \(\vec{P}\) and angular momenta \(\vec{M}\) are the complete
linear independent set of additive first integrals.

Let us present now our derivation of Gibbs distribution (from the
cluster property). This derivation is not rigorous too but we hope
it is of some interest. Note that this derivation uses some basic
ideas of Haag-Ruelle scattering theory [7].

But at first let us give some previous definitions.

\textbf{Definition 1 .}Let \(S( \mathbb{R}^3)\) be a Schwatrz space
of test functions (infinitely-differentiable functions decaying at
infinity faster than any inverse polynomial with all its
derivatives). The algebra of canonical commutative relations
\(\mathcal{C}\) is an unital algebra generated by symbols \(a^+(f)\)
and \(a(f)\) \(f \in S( \mathbb{R}^3)\) satisfying the following
canonical commutative relations:

a) \(a^+(f)\) is a linear functional of \(f\),

b) \(a(f)\) is an antilinear functional of \(f\),

\begin{eqnarray}
{[a(f),a(g)]}={[a^+(f),a^+(f)]}=0,\nonumber\\
{[a(f),a^+(g)]}=\langle f,g\rangle,\nonumber
\end{eqnarray}
where \( \langle f,g\rangle\) is a standard scalar product in
\(L^2(\mathbb{R}^3)\),
\begin{eqnarray}
\langle f,g\rangle:=\int f^*(x)g(x) d^3 x\nonumber
\end{eqnarray}
\textbf{Remark.} We will widely use generalized "elements" of the
algebra of canonical commutative relations \(a(k)\), \(a^+(k)\)
defined according to the following relations
\begin{eqnarray}
a(f)=\int  a(k)f(k)d^3k,\nonumber\\
a^+(f)=\int  a^+(k) f^+(k)d^3k.
\end{eqnarray}

\(a(k)\), \(a^+(k')\) are called the annihilation-creation operators
and satisfy to the following canonical commutative relations:
\begin{eqnarray}
{[a^+(k),a^+(k')]}={[a(k),a(k')]}=0,\nonumber\\
{[a(k),a(k')]}=\delta(k-k').
\end{eqnarray}

\textbf{Definition 2.} The field operators \(\Psi(x)\),
\(\Psi^+(x)\) are defined as follows
\begin{eqnarray}
\Psi(x)=\frac{1}{(2\pi)^{\frac{3}{2}}} \int
a(k)e^{ikx}d^3k,\nonumber\\
\Psi^+(x)=\frac{1}{(2\pi)^{\frac{3}{2}}} \int a^+(k)e^{ikx}d^3k
\end{eqnarray}
The rigorous definition of \(\Psi(x)\) and \(\Psi^+(x)\) could be
obtained from this definition by using the notion of the Fourier
transform of distributions.

\textbf{Definition 3.} Let \(\rho\) be a state on the algebra of
canonical commutative relations (the algebra generated by smoothed
secondary quantized wave functions) (CCR-algebra). We say that
\(\rho\) is a Gauss state if we can calculate its values at elements
of CCR-algebra by using the Wick (Bloch --- De Dominicis) theorem
through pair correlations.

Now let us recall some notions connected with nonequilibrium
(Keldysh) diagram technique. Let \(\rho\) be an arbitrary state on
the algebra of canonical commutative relations. Let us introduce the
Green functions for the system (corresponding to this state).

\begin{eqnarray}
\rho(T(\Psi^\pm_H(t_1, x_1),...,\Psi^\pm_H(t_n, x_n))).\nonumber
\end{eqnarray}

Here \(t_1,...,t_n\) are times, symbol \(H\) near \(\Psi^{\pm}\)
means here that \(\Psi^\pm_H\) are operators in Heizenberg
representation and the symbol \(T\) is a symbol of chronological
ordering.

We will consider the system describing by the following Hamiltonian
\begin{eqnarray}
H=H_0+\lambda V,
\end{eqnarray}
where \(H_0\) is a free Hamiltonian
\begin{eqnarray}
H_0=\int \omega(k)a^+(k)a(k)d^3 k,
\end{eqnarray}
\begin{eqnarray}
\omega(k)=\frac{k^2}{2}.
\end{eqnarray}
\(\lambda \in \mathbb{R}\) is a coupling constant and
\begin{eqnarray}
V=\frac{1}{2}\int d^3x d^3
x'\Psi^+(x)\Psi^+(x')\mathcal{V}(x-x')\Psi(x')\Psi(x),
\end{eqnarray}
where \(\mathcal{V}(x)\) is an arbitrary test function from
\(S(\mathbb{R}^3)\). In nonequilibrium diagram technique we require
the following representation for the Green functions:
\begin{eqnarray}
\rho(T(\Psi^\pm_H(t_1,x_1),...,\Psi^\pm_H(t_n,x_n)))=\nonumber\\
\rho_0(S^{-1}T(\Psi^\pm_0(t_1,x_1),...,\Psi^\pm_0(t_n,x_n)S)).\label{Green}
\end{eqnarray}
The symbol \(0\) near \(\Psi^{\pm}\) means here that
\(\Psi^{\pm}_0\) are operators in the Dirac representation
(representation of interaction). The \(S\)-matrix has the form
\begin{eqnarray}
S=T\rm exp \mit{(-i\int
\limits_{-\infty}^{+\infty}V_0(t)dt)},\;\nonumber
\end{eqnarray}
and
\begin{eqnarray}
S^{-1}=\tilde{T}\rm exp \mit{(i\int
\limits_{-\infty}^{+\infty}V_0(t)dt)}.\nonumber
\end{eqnarray}
Here \(T\) and \(\tilde{T}\) are symbols of the chronological and
the antichronological ordering respectively. \(\rho_0\) is some
Gauss state defined by density function \(n(k)\) as follows
\begin{eqnarray}
\rho_0(a^+(k')a^+(k))=\rho_0(a(k)a(k'))=0,\nonumber\\
\rho_0(a^+(k)a(k'))=n(k)\delta(k-k').
\end{eqnarray}
The state \(\rho_0\) is called the asymptotical state.

Let us recall the basic elements of nonequilibrium diagram
technique. The vertices coming from \(T\)-exponent are marked by
symbol \(-\). The vertices coming from \(\tilde{T}\)-exponent are
marked by symbol \(+\). There exist four types of propagators
\begin{eqnarray}
G_0^{+-}(t_1-t_2,x_1-x_2):=\rho_0(\Psi(t_1,x_1)\Psi^{+}(t_2,x_2)),\nonumber\\
G_0^{-+}(t_1-t_2,x_1-x_2):=\rho_0(\Psi^+(t_2,x_2)\Psi(t_1,x_1)),\nonumber\\
G_0^{--}(t_1-t_2,x_1-x_2):=\rho_0(T(\Psi(t_1,x_1)\Psi^+(t_2,x_2))),\nonumber\\
G_0^{++}(t_1-t_2,x_1-x_2):=\rho_0(\tilde{T}(\Psi(t_1,x_1)\Psi^+(t_2,x_2))).
\nonumber
\end{eqnarray}
Let us write the table of propagators
\begin{eqnarray}
G_0^{+-}(t,x)=\int \frac{d^4k}{(2\pi)^4}(2\pi)\delta(\omega-\omega(k))(1+n(k))e^{-i(\omega t-kx)},\nonumber\\
G_0^{-+}(t,x)=\int \frac{d^4k}{(2\pi)^4}(2\pi)\delta(\omega-\omega(k))n(k)e^{-i(\omega t-kx)},\nonumber\\
G_0^{--}(t,x)=i\int
\frac{d^4k}{(2\pi)^4}\{\frac{1+n(k)}{\omega-\omega(k)+i0}
-\frac{n(k)}{\omega-\omega(k)-i0}\}e^{-i(\omega t-kx)},\nonumber\\
G_0^{++}(t,x)=i\int
\frac{d^4k}{(2\pi)^4}\{\frac{n(k)}{\omega-\omega(k)+i0}
-\frac{1+n(k)}{\omega-\omega(k)-i0}\}e^{-i(\omega t-kx)}.\nonumber
\end{eqnarray}
It has been shown in [1] that (usually) there exists divergences in
Keldysh diagram technique. A typical example of divergent diagrams
is pictured at fig.1

\begin{picture}(300,100)
\put(10,80){fig. 1} \put(260,50){\vector(-1,0){45}}
\put(200,50){\oval(30,20)} \put(185,50){\vector(-1,0){15}}
\put(130,50){\vector(-1,0){15}} \put(85,50){\vector(-1,0){45}}
\put(100,50){\oval(30,20)} \multiput(162,50)(-11,0){3}{\circle*{4}}
\put(260,45){\line(-1,0){6}} \put(218,41){+}
\put(224,39){\line(-1,0){6}} \put(176,41){+}
\put(176,39){\line(1,0){6}} \put(118,41){+}
\put(124,39){\line(-1,0){6}}
 \put(76,41){+}
\put(76,39){\line(1,0){6}} \put(40,45){\line(1,0){6}}
\end{picture}

The ovals represent the sum of one-particle irreducible diagrams.
These diagrams are called chain diagrams. Let us suppose that all
divergences of self-energy parts (ovals) are subtracted. The
divergences arise from the fact that singular supports of
propagators coincide.

For all \(t_1,t_2 \in \mathbb{R}\) let us define the evolution
operator \(S(t_1,t_2)\) as follows:
\begin{eqnarray}
S(t_1,t_2)=e^{it_1H_0}e^{-i(t_1-t_2)H}e^{-itH_0}.
\end{eqnarray}
Note that, for example, that
\begin{eqnarray}
S(0,-\infty)=T\rm exp \mit (-i\int \limits_{-\infty}^{0}
V_0(t)dt), \nonumber\\
S(+\infty, 0)=T\rm exp \mit (-i\int \limits_{0}^{+\infty} V_0(t)dt)
\end{eqnarray}
One can prove that the following lemma holds.

\textbf{Lemma 1.} \(\forall t \in \mathbb{R}\) the following
equalities holds:
\begin{eqnarray}
e^{-itH}S(0,-\infty)=S(0,-\infty)e^{-itH_0}\nonumber\\
\end{eqnarray}
and
\begin{eqnarray}
S(+\infty,0)e^{-itH}=e^{-itH_0}S(+\infty,0).
\end{eqnarray}
Now let us consider the equilibrium Green functions defined as
follows
\begin{eqnarray}
\frac{1}{Z} \rm tr \mit
(T(\Psi^{\pm}_H(t_1,x_1)...\Psi^{\pm}_H(t_n,x_n))e^{-\frac{H-\mu
N}{T}}),
\end{eqnarray}
where
\begin{eqnarray}
Z=\rm tr \mit(e^{-\frac{H-\mu N}{T}}),\label{statsum}
\end{eqnarray}
\(T>0\) is a temperature, \(\mu \in \mathbb{R}\) is a chemical
potential and \(N\) is a number particle operator
\begin{eqnarray}
N=\int  a^+(k)a(k)d^3 k
\end{eqnarray}
In other words we put in (\ref{Green})
\begin{eqnarray}
\rho(\cdot)=\frac{1}{Z} \rm tr \mit ((\cdot)e^{-\frac{-H-\mu
T}{T}}).
\end{eqnarray}
It follows from the lemma 1 that the following lemma holds.

\textbf{Lemma 2.}
\begin{eqnarray}
\frac{1}{Z} \rm tr \mit
(T(\Psi^{\pm}_H(t_1,x_1)...\Psi^{\pm}_H(t_n,x_n))e^{-\frac{H-\mu
N}{T}})\nonumber\\
=\rho_0(S^{-1}T(\Psi^\pm_0(t_1,x_1),...,\Psi^\pm_0(t_n,x_n)S)),
\end{eqnarray}
where \(\rho_0\) is a Gauss state on the CCR-algebra defined as
usual by its density function \(n(k)\) of the form:
\begin{eqnarray}
n(k)=\frac{1}{e^{\frac{\omega(k)-\mu}{T}}-1}
\end{eqnarray}
It follows from physical reasonings that the equilibrium Green
functions does not contain divergences. So we have the following
lemma:

\textbf{Lemma 3.} If the asymptotical state \(\rho_0\) in Keldysh
diagram technique corresponds to the density function \(n(k)\) of
Bose-Einstein form
\begin{eqnarray}
n(k)=\frac{1}{e^{\frac{\omega(k)-\mu}{T}}-1}
\end{eqnarray}
then the Keldysh diagram technique does not contain divergences and
corresponding green function are equilibrium
\begin{eqnarray}
\rho(T(\Psi^\pm_H(t_1,x_1),...,\Psi^\pm_H(t_n,x_n)))\nonumber\\
=\frac{1}{Z} \rm tr \mit
(T(\Psi^{\pm}_H(t_1,x_1)...\Psi^{\pm}_H(t_n,x_n))e^{-\frac{H-\mu
N}{T}}),
\end{eqnarray}
Now let us prove the fact that if the Keldysh diagram technique does
not contain divergences then the corresponding Green functions are
equilibrium, i.e. the asymptotical state \(\rho_0\) has
Bose-Einstein form. In order to do this let us analyze the
divergences of Keldysh diagram technique in lowest possible order in
\(\lambda\). Such divergences may come only from the diagrams
pictured at fig. 2.

\begin{picture}(200,100)
\put(10,80){fig. 2} \put(160,50){\vector(-1,0){45}}
\put(100,50){\oval(30,20)} \put(85,50){\vector(-1,0){45}}
\put(160,45){\line(-1,0){6}} \put(118,41){+}
\put(124,39){\line(-1,0){6}} \put(76,41){+}
\put(76,39){\line(1,0){6}} \put(40,45){\line(1,0){6}}
\end{picture}
where ovals means here the sum of all self-energy diagrams of lowest
possible order in \(\lambda\). It has been shown in [1] that these
diagrams contain divergences for density function \(n(k)\) of
general form. It has been also shown in [1] that these diagrams
could be subtracted by the following renormalization of the
asymptotical state

\begin{eqnarray}
\rho_0(\cdot)\rightarrow \frac{1}{Z}\rho_0(e^{-\int
\limits_{-\infty}^{+\infty} h_0(t)dt}(\cdot)),\nonumber
\end{eqnarray}
where
\begin{eqnarray}
h=\int h(k)a^+(k)a(k)d^3k,\nonumber
\end{eqnarray}
\(h(k)\) is a real-valued function,
\begin{eqnarray}
h_0(t):=e^{itH_0}he^{-itH_0}
\end{eqnarray}
and
\begin{eqnarray}
Z=\rho_0(e^{-\int \limits_{-\infty}^{+\infty} h(t)dt})\nonumber
\end{eqnarray}
for suitable \(h=\int h(k)a^+(k)a(k)d^3k\).

It also could be extracted from [1] that the divergences in
considered diagrams exists if and only if \(h\neq 0\). But it has
been also shown in [1] that
\begin{eqnarray}
h(p)=\frac{1+2n(p)}{2n(p)(1+n(p))} St(p),
\end{eqnarray}
where \(St(p)\) is a collision integral (in lowest possible order of
perturbation theory) and \(St(p)\equiv 0\) if and only if \(n(p)\)
has a Bose-Einstein form (see for example [8]). So we have proved
the following

\textbf{Lemma 4.} If Keldysh diagram technique does not contain
divergences the the asimptotical state has a Bose-Einstein form
\begin{eqnarray}
\rho_0(a^+(k')a^+(k))=\rho_0(a(k)a(k'))=0,\nonumber\\
\rho_0(a^+(k)a(k'))=n(k)\delta(k-k'), \nonumber\\
n(k)=\frac{1}{e^{\frac{\omega(k)-\mu}{T}}-1}
\end{eqnarray}
and the corresponding Green function are equilibrium.

Now let \(\langle\cdot\rangle\) be some stationary translation
invariant state on the CCR-algebra, commuting with the number
particle operator. Suppose that \(\langle\cdot\rangle\) satisfy to
the asymptotic property of (enough fast) space factorization of
correlations. Put by definition
\begin{eqnarray}
W=S(0,-\infty).
\end{eqnarray}
Let \(\langle\cdot\rangle_0\) be a state on the CCR-algebra such
that
\begin{eqnarray}
\langle\cdot\rangle=\langle W^{-1}(\cdot)W\rangle_0.
\end{eqnarray}
One has
\begin{eqnarray}
\langle\cdot\rangle_0=\langle W(\cdot)W^{-1}\rangle.
\end{eqnarray}
Let us formulate now our main lemma.

\textbf{Lemma 5.} If the state \(\langle\cdot\rangle\) satisfies to
the property of asymptotic (enough fast) factorization of
correlations then \(\langle\cdot\rangle_0\) is finite (does not
contain divergences) and Gauss.

\textbf{Proof.} We have
\begin{eqnarray}
W=\lim \limits_{t\rightarrow -\infty} e^{+itH}e^{-itH_0}
\end{eqnarray}
and
\begin{eqnarray}
W^{-1}=\lim \limits_{t\rightarrow -\infty} e^{+itH_0}e^{-itH}.
\end{eqnarray}
These two equations could be simply established in the sense of
formal series of perturbation theory and this enough for our aims.

Therefore we have
\begin{eqnarray}
\langle\cdot\rangle_0=\langle W(\cdot)W^{-1}\rangle\nonumber\\
=\lim \limits_{t_1\rightarrow -\infty,\;t_2\rightarrow -\infty}
\langle
e^{+it_1H}e^{-it_1H_0}(\cdot)e^{+it_2H_0}e^{-it_2H}\rangle\nonumber\\
=\lim \limits_{t\rightarrow -\infty} \langle
e^{+itH}e^{-itH_0}(\cdot)e^{+itH_0}e^{-itH}\rangle\nonumber\\
=\lim \limits_{t\rightarrow -\infty} \langle
e^{-itH_0}(\cdot)e^{+itH_0}\rangle.
\end{eqnarray}
Last equality takes place because the state \(\langle\cdot\rangle\)
is a stationary state. In result
\begin{eqnarray}
\langle\cdot\rangle_0=\lim \limits_{t\rightarrow -\infty} \langle
e^{-itH_0}(\cdot)e^{+itH_0}\rangle.
\end{eqnarray}
Let us now introduce so-called Wightman functions (distributions)
\(W_{n,t}(x_1,\sigma_1,...,x_n,\sigma_n)\), \(t \in\mathbb{R}\)
\(x_1,...,x_n \in \mathbb{R}^3\), \(\sigma_1,...,\sigma_n \in
\{+,-\}\) as follows:
\begin{eqnarray}
W_{n,t}(x_1,\sigma_1,...,x_n,\sigma_n):=\langle
e^{itH_0}(\Psi^{\sigma_1}(x_1)...\Psi^{\sigma_n}(x_n))
e^{-itH_0}\rangle.
\end{eqnarray}
Let us also introduce so-called truncated Wightman functions
according \(W^T_t\) to following relations:
\begin{eqnarray}
W_{n,t}(x_1,\sigma_1,...,x_n,\sigma_n)=\sum \limits_{\pi \in
\mathcal{P}} \prod \limits_{P \in
\pi}W^T_{|P|,t}(x_{i_1^P},\sigma_{i_1^P},...,x_{i_{|P|}^P},\sigma_{i_{|P|}^P}).
\label{cluster}
\end{eqnarray}
Here \(|A|\) is a number of elements of finite set \(A\),
\(\mathcal{P}\) is a set of all decompositions
\(\pi=\{P_1,...,P_{|\pi|}\}\) of the set \(\{1,...,n\}\) into
disjoint union of sets \(P_1,...,P_{|\pi|}\). For each
\(k=1,...,|\pi|\) we put by definition
\(P_k=\{i_1^{P_k},...,i_{|P_k|}^{P_k}\}\),
\(i_1^{P_k}<i_2^{P_k}<...<i_{|P_k|}^{P_k}\).

Note that the property of enough fast asymptotic space factorization
of correlations by definition means that truncated Wightman
functions \(W_{n,t}^T(x_1,\sigma_1,...,x_n,\sigma_n)\) tends to zero
enough fast if their relative arguments
\(\xi_1=x_2-x_1,...\xi_{n-1}=x_n-x_1\) tends to infinity.

Now let us introduce the Fourier transformations of Wightman
functions as follows:
\begin{eqnarray}
\tilde{W}_{n,t}(p_1,\sigma_1,...,p_n,\sigma_n):\nonumber\\
=\int d^3x_1,...,d^3 x_n
e^{i(\sigma_1p_1x_1+...+\sigma_np_nx_n)}W_{n,t}(x_1,\sigma_1,...,x_n,\sigma_n)
\end{eqnarray}
and
\begin{eqnarray}
\tilde{W}_{n,t}^T(p_1,\sigma_1,...,p_n,\sigma_n):\nonumber\\
=\int d^3x_1,...,d^3 x_n
e^{i(\sigma_1p_1x_1+...+\sigma_np_nx_n)}W_{n,t}^T(x_1,\sigma_1,...,x_n,\sigma_n).
\end{eqnarray}
one can prove that
\begin{eqnarray}
\tilde{W}_{n,t}(p_1,\sigma_1,...,p_n,\sigma_n)=\sum \limits_{\pi \in
\mathcal{P}} \prod \limits_{P \in
\pi}\tilde{W}^T_{|P|,t}(p_{i_1^P},\sigma_{i_1^P},...,p_{i_{|P|}^P},\sigma_{i_{|P|}^P}),
\end{eqnarray}
where notations are such that as in (\ref{cluster}).

The truncated Wightman functions
\(W_{n,t}^T(x_1,\sigma_1,...,x_n,\sigma_n)\) are
translation-invariant and the function of (enough) fast decay on
relative arguments. Therefore their Fourier transforms
\(\tilde{W}_{n,t}^T\) could be represented as follows:
\begin{eqnarray}
\tilde{W}_{n,t}(p_1,\sigma_1,...,p_n,\sigma_n)=\delta(p_1\sigma_1+...+p_n\sigma_n)
\widetilde{\mathcal{W}}_{n,t}^T(p_1,\sigma_1,...,p_n,\sigma_n),
\end{eqnarray}
where \(\widetilde{\mathcal{W}}_{n,t}^T\) is an enough smooth
function of at most polynomial increment at infinity (with enough
large number of derivatives).

It is obvious that \(\forall n=1,2,...\) and \(\forall t \in
\mathbb{R}\)
\begin{eqnarray}
\tilde{W}_{n,t}^T(p_1,\sigma_1,...,p_n,\sigma_n)=e^{i(\sigma_1\omega(p_1)+...+\sigma_n\omega(p_n))t}\tilde{W}_{n,0}^T(p_1,\sigma_1,...,p_n,\sigma_n).
\end{eqnarray}
Now, let us prove that for each \(n=3,4,...\)
\(\tilde{W}_{n,t}(p_1,\sigma_1,...,p_n,\sigma_n)\) tends to zero as
\(t\rightarrow -\infty\) in the sense of distributions. This means
that for each test function \(f(p_1,...,p_n) \in
S(\mathbb{R}^{3n})\)
\begin{eqnarray}
\int d^3p_1...d^3p_n
\tilde{W}_{n,t}^T(p_1,\sigma_1,...,p_n,\sigma_n)f(p_1,...,p_n)\rightarrow
0
\end{eqnarray}
as \(t\rightarrow -\infty\). It is obvious from remarks above that
to prove this fact it is enough to prove that for each enough smooth
function \(g(x_1,...,x_n)\) of enough fast decaying at infinity
\begin{eqnarray}
\int d^3p_1...d^3p_n
e^{i(\sigma_1\omega(p_1)+...+\sigma_n\omega(p_n))t}\delta(p_1\sigma_1+...+p_n\sigma_n)
g(p_1,...,p_n)\rightarrow 0 \label{StatPhase}
\end{eqnarray}
as \(t\rightarrow -\infty\).

Put by definition
\(L=\{(p_1,...,p_n)|\sigma_1p_1+...\sigma_np_n=0\}\). Note that for
 each \(n=3,4,...\) \(\sigma_1\omega_1(p)+...+\sigma_n\omega_n(p)\) is not identically
 equal to zero on \(L\). This fact could pe proved very simply but
 we omit this proof. So the limit equality (\ref{StatPhase}) could
 be directly obtained by using the stationary phase method.

 So, for each \(n=3,4,...\) \(\tilde{W}_{n,t}(p_1,\sigma_1,...,p_n,\sigma_n)\rightarrow
 0\) as \(t\rightarrow -\infty\) in the sense of distributions. From
 other hand, it is obvious that \(W^T_{2,t}=W^T_{2,0}\). In result
 \(\forall t \in \mathbb{R}\)
\begin{eqnarray}
{W}_{n,t}(x_1,\sigma_1,...,x_n,\sigma_n)=0
\end{eqnarray}
if \(n\) is odd and
\begin{eqnarray}
\lim \limits_{t\rightarrow
-\infty}{W}_{n,t}(x_1,\sigma_1,...,x_n,\sigma_n)=\sum
\limits_{\tau\in \mathcal{T}} \prod \limits_{T \in \tau}
W_{2,0}(x_{i_T},\sigma_{i_T},x_{j_T},\sigma_{j_T}),
\end{eqnarray}
if \(n\) is even (in the sense of distributions). Here
\(\mathcal{T}\) is a set of all decomposition
\(\tau=\{T_1,...,T_{\frac{n}{2}}\}\) of the set \(\{1,2,...,m\}\)
into disjoint union of pairs \(\tau_k=\{i_{\tau_k},j_{\tau_k}\}\),
\(i_{\tau_k}<j_{\tau_k}\), \(k=1,2,...,\frac{n}{2}\).

But let us recall that
\begin{eqnarray}
\lim \limits_{t\rightarrow -\infty}
W_{n,t}(x_1,\sigma_1,...,x_n,\sigma_n)=\langle\Psi^{\sigma_1}(x_1)...Psi^{\sigma_n}(x_n)\rangle_0.
\end{eqnarray}
in the sense of distributions. Therefore \(\langle\cdot\rangle_0\)
is finite and Gauss. Now let us formulate main theorem of present
section

\textbf{Theorem 2.} let \(\langle\cdot\rangle\) be some stationary
translation invariant state on the CCR-algebra, commuting with the
number particle operator. Suppose that \(\langle\cdot\rangle\)
satisfy to the asymptotic property of (enough fast) space
factorization of correlations. Then
\begin{eqnarray}
\rho(\cdot)=\frac{1}{Z} \rm tr \mit ((\cdot)e^{-\frac{-H-\mu N}{T}})
\end{eqnarray}
for some \(T>0\) and \(\mu \in \mathbb{R}\). \(Z\) is a statistical
sum defined by equation (\ref{statsum}). In other words the state
\(\langle\cdot\rangle\) is a Gibbsian state.

\textbf{Proof.} It follows from lemma 5 that
\begin{eqnarray}
\langle\cdot\rangle\nonumber\\
=\langle S^{-1}(0,-\infty)(\cdot)S(0,-\infty)\rangle_0.
\end{eqnarray}
This implies that
\begin{eqnarray}
\langle T\Psi^{\pm}_H(t_1,x_1)...\Psi^{\pm}_H(t_n,x_n)\rangle\nonumber\\
=\langle
S^{-1}T\Psi^\pm_0(t_1,x_1),...,\Psi^\pm_0(t_n,x_n)S\rangle_0.
\end{eqnarray}
But one can prove (in the sense of formal series of perturbation
theory) that the fact that \(\langle\cdot\rangle\) finite implies
that \(\langle
T\Psi^{\pm}_H(t_1,x_1)...\Psi^{\pm}_H(t_n,x_n)\rangle\) is finite.
Therefore the Keldysh diagram technique for \(\langle
T\Psi^{\pm}_H(t_1,x_1)...\Psi^{\pm}_H(t_n,x_n)\rangle\) does not
contain divergences. Therefore lemma 4 implies that
\begin{eqnarray}
\langle a^+(k)a^+(k')\rangle_0=\langle a(k)a(k')\rangle_0=0,\nonumber\\
\langle
a(k)a^+(k')\rangle_0=\delta(k-k')\frac{1}{e^{\frac{\omega(k)-\mu}{T}}-1}
\end{eqnarray}
for some \(T>0\) and \(\mu \in \mathbb{R}\). And it follows from
lemma 2 that
\begin{eqnarray}
\langle\cdot\rangle=\frac{1}{Z} \rm tr \mit ((\cdot)e^{-\frac{-H-\mu
T}{T}}).
\end{eqnarray}
Therefore theorem is proved.

\textbf{Remark.} Note that we have supposed the following when we
have proved the theorem 2 above. If the state
\(\langle\cdot\rangle\) is a finite state (does not contain the
divergences) then the divergences does not exist in each order in
coupling constant \(\lambda\) of the power decomposition of the
state \(\langle\cdot\rangle\) according to perturbation theory. We
have not been prove this statement but it is essential to suppose
that this statement holds for all states of the physical interest.

\section{The Bardeen --- Cooper --- Schrieffer model on the language of generalized thermodynamics.}
In this section we show how to formulate the Bardeen --- Cooper
--- Schrieffer model of suoercondactivity on the language of the generalized
thermodynamics. But at first let us recall several results on some
model Hamiltonian in superconductive theory which belong to
Bogoliubov --- Zubarev --- Tserkovikov [5,9].

Let us consider a model dynamical system described by the following
Hamiltonian:
\begin{eqnarray}
H=\sum \limits_{f} T(f)a^+_fa-\frac{1}{2V} \sum \limits_{f',f}
\lambda(f)\lambda(f') a^+_fa^+_{-f}a_{-f'}a_{f'}.\label{Hamiltonian}
\end{eqnarray}
Here we use the following notations:
\begin{eqnarray}
f=(p,s),\;-f=(-p,-s).
\end{eqnarray}
\(p\) is a particle's momentum, \(s=\pm 1\),
\begin{eqnarray}
T(f)=\frac{p^2}{2m}-\mu,\; \mu>0.
\end{eqnarray}
\begin{eqnarray}
\lambda(f)=J\varepsilon(f)\; \rm if\mit\;
|\frac{p^2}{2}-\mu|<\Delta,\nonumber\\
\lambda(f)=0\; \rm if \mit\; |\frac{p^2}{2}-\mu|>\Delta,\nonumber\\
\varepsilon((p,s))=s,
\end{eqnarray}

\(a_f,\;a^+_f\) are usual fermion annihilation-creation operators.

We will also consider the Hamiltonian \(H\) with some additional
terms which are the sources of creation and annihilation of pairs.
\begin{eqnarray}
H_\nu=H-\frac{\nu}{2}\sum \limits_f
\lambda(f)\{a_{-f}a_f+a^+_fa^+_{-f}\},\nonumber\\
\nu>0.
\end{eqnarray}
Let \(C\) be an arbitrary complex number. Let us represent \(H\) as
follows:
\begin{eqnarray}
H=H_0(C)+H_1(C),
\end{eqnarray}
where
\begin{eqnarray}
H_0(C)=\sum \limits_f T(f)a^+_fa_f-\frac{1}{2}\sum \limits_f
\lambda(f)\{C^{\ast}a_{-f}a_f+Ca^+_fa^+_{-f}\}+\frac{|C|^2}{2}V
\end{eqnarray}
and
\begin{eqnarray}
H_1(C)=\frac{V}{2}(\frac{1}{V}\sum
\limits_f\lambda(f)a^+_fa^+_{-f}-C^{\ast})(\frac{1}{V}\sum
\limits_f\lambda(f)a_{-f}a_{f}-C)
\end{eqnarray}
Note that \(H_0(C)\) is a quadratic form on fermion
annihilation-creation operators. So it may be diagonalized by mens
of canonical fermion Bogoliubov transformation. More precisely:
\begin{eqnarray}
H_0(C)=\sum \limits_f
\sqrt{\lambda^+(f)|C|^2+T^2(f)}(a^+_fu_f+a_{-f}v^{\ast}_f)(a_fu_f+a^+_{-f}v_f)+\nonumber\\
+\frac{V}{2}(|C|^2-\frac{1}{V}\sum
\limits_f[\sqrt{\lambda^2(f)|C|^2+T^2(f)}-T(f)]),
\end{eqnarray}
where
\begin{eqnarray}
u_f=\frac{1}{\sqrt{2}}(1+\frac{T(f)}{\sqrt{\lambda^2(f)|C|^2+T^2(f)}})^{{1}/{2}},
\end{eqnarray}
\begin{eqnarray}
v_f=-\frac{\varepsilon(f)}{\sqrt{2}}(1-\frac{T(f)}{\sqrt{\lambda^2(f)|C|^2+T^2(f)}})^{1/2}\frac{C}{|C|}
\end{eqnarray}
Note that \(u(-f)=u(f)\) and \(v(-f)=-v(f)\). Function \(u\) is a
real-valued function and function \(v\) is a complex-valued
function, and \(u^2+|v|^2=1\). These implies that the operators
\begin{eqnarray}
\alpha_f:=a_fu_f+a^+_{-f}v_f, \nonumber\\
\alpha^+_f=a^+_fu_f+a_{-f}v^{\ast}_f
\end{eqnarray}
are the operators of the fermion type.

Put by definition:
\begin{eqnarray}
J(C):=\frac{1}{2V}\sum \limits_f \lambda(f)a^+_fa_{-f}-C^{\ast}.
\end{eqnarray}
Now the expression for \(H_1(C)\) could be rewritten as follows:
\begin{eqnarray}
H_1(C)=-\frac{V}{2}J^{\ast}(C)J(C).
\end{eqnarray}
Let \(\Gamma\) be an arbitrary (low bounded) Hamiltonian. Denote by
\(\langle\cdot\rangle_{\Gamma,T}\)the averaging with respect to the
Gibbs distribution corresponding to the Hamiltonian \(\Gamma\) and
the temperature \(T\):
\begin{eqnarray}
\langle\cdot\rangle_{\Gamma,T}:=\frac{\rm tr
\mit((\cdot)e^{-\frac{\Gamma}{T}})}{\rm tr \mit
(e^{-\frac{\Gamma}{T}})}.
\end{eqnarray}
 Let us consider the following equation:
\begin{eqnarray}
\langle J(C)\rangle_{H_0(C),T}=0 \label{Compens}
\end{eqnarray}
on the variable \(C\). This equation is called the equation of
compensation. Note that \(H_0(C)\) is a quadratic form on
creation-annihilation operators. Therefore the equation of
compensation could be rewritten explicitly as follows [5]:
\begin{eqnarray}
1=\frac{1}{2V}\sum \limits_f
\frac{\lambda^2(f)}{\sqrt{\lambda^2(f)|C|^2+T^2(f)}}\rm th \mit
\{\frac{\sqrt{\lambda^2(f)|C|^2+T^2(f)}}{2T}\}.\label{II}
\end{eqnarray}
In the limit \(V \rightarrow \infty\) this equation will take the
form:
\begin{eqnarray}
1=\frac{1}{(2 \pi)^2} \int
\frac{\lambda(p)d^3p}{\sqrt{\lambda^2(p)|C|^2+T^2(p)}} \rm th \mit
\{\frac{\sqrt{\lambda^2(f)|C|^2+T^2(p)}}{2T}\}.\label{III}
\end{eqnarray}
It is well-known fact that last equation has a non-zero solution for
all temperature \(T<T_{cr}\) for some critical temperatures
\(T<T_{cr}\) for some temperature \(T_{cr}\). We will consider only
the case when \(T<T_{cr}\). Let \(C_0\) be an arbitrary solution of
the equation (\ref{Compens}) or equation (\ref{II}). Note that if
\(C_0\) is a solution of equation (\ref{II}) then \(C_0 e^{i\phi}\)
is a solution of equation (\ref{II}) too for an arbitrary real
number \(\varphi\). Let \(C'_0\) is a positive solution of the
equation (\ref{I}) or equation (\ref{II}). Now let us  formulate
main results of Bogoliubov, Zubarev, Tserkovnikov [9] concerned with
the system under consideration.

Let \(\Gamma\) be a Hamiltonian of some system contained in some
volume V). Denote by \(F_\Gamma(V,T)\) the free (Gibbsian) energy of
this system corresponding to the temperature \(T\). Then the
following equalities hold
\begin{eqnarray}
\lim \limits_{V\rightarrow\infty} \frac{F_H(V,T)}{V}=\lim
\limits_{\nu\rightarrow \infty,\;\nu>0} \lim \limits_{V\rightarrow
\infty}\frac{F_{H_\nu}(V,T)}{V}=\lim
\limits_{V\rightarrow\infty}\frac{F_{H(C_0)}(V,T)}{V}. \label{IV}
\end{eqnarray}
Now, let us consider quasi-averages
\begin{eqnarray}
\prec...a^+_{f_j}...a_{f_s}...\succ_{H,T}:=\nonumber\\
=\lim \limits_{\nu\rightarrow 0,\;\nu>0} \lim \limits_{V\rightarrow
\infty} \langle...a^+_{f_j}...a_{f_s}...\rangle_{H_\nu,T}
\end{eqnarray}
Here \(\langle...a^+_{f_j}...a_{f_s}...\rangle_{H_\nu,T}\) are
fermion creation-annihilation operators in Heisenberg
representation. The following equation holds [9,5]:
\begin{eqnarray}
\prec...a^+_{f_j}...a_{f_s}...\succ_{H,T}\nonumber\\
= \lim \limits_{V\rightarrow \infty}
\langle...a^+_{f_j}...a_{f_s}...\rangle_{H(C'_0),T}.
\end{eqnarray} Now let us consider the following Hamiltonian
\(H_\nu^\varphi\) instead of \(H_\nu\):
\begin{eqnarray}
H_\nu^\varphi=-\frac{\nu}{2}\sum \limits_f
\lambda(f)\{a_{-f}a_fe^{i\varphi}+a^+_fa^+_{-f}e^{-i\varphi}\},
\end{eqnarray}
where \(\varphi\) is an arbitrary. We will have again the equality
(\ref{IV}), or more precisely:
\begin{eqnarray}
\lim \limits_{V\rightarrow\infty} \frac{F_H(V,T)}{V}=\lim
\limits_{\nu\rightarrow \infty,\;\nu>0} \lim \limits_{V\rightarrow
\infty}\frac{F_{H_\nu^\varphi}(V,T)}{V}=\lim
\limits_{V\rightarrow\infty}\frac{F_{H(C_0)}(V,T)}{V}.
\end{eqnarray}
where we put now
\begin{eqnarray}
\prec...a^+_{f_j}...a_{f_s}...\succ_{H,T}\nonumber\\
=\lim \limits_{\nu\rightarrow 0,\;\nu>0} \lim \limits_{V\rightarrow
\infty} \langle...a^+_{f_j}...a_{f_s}...\rangle_{H_\nu^\varphi,T}
\end{eqnarray}
and
\begin{eqnarray}
C_0^\varphi:=C_0e^{i\varphi}.
\end{eqnarray}
It is necessary to note that the equation of compensation (\ref{II})
is equal to a condition of the minimum of free energy
\(F_{H(C_0)}(V,T)\) on \(C\) at fixed energy \(T\)) (see [5]).
Recall that \(F_{H_0(C)}\) is a free energy of the dynamical system
which is described by the Hamiltonian \(H_0(C)\) and contained in
volume \(V\) at the temperature \(T\). But the condition that
\(C_0\) corresponds to the minimum of free energy \(F_{H_0(C_0)}\)
is equivalent to the following condition: the entropy of our system
\(S_{H_0(C)}\) achieve the maximum at the point \(C=C_0\) (as a
function of \(C\)) at fixed energy \(E\). Last remark admit us to
suppose that thermodynamics of our model could be described in terms
of our generalized thermodynamics and the parameter \(C\) here plays
the role of the observable values of commuting integrals of the
system.

Let us recall that  one of the basics principles of Generalized
thermodynamics states the following. The observable values of of the
commuting first integrals (from microcanonical distribution) which
may be realized in nature must corresponds to the maximum of the
entropy (under fixed energy).

Let us start to reformulate thermodynamics of our model in terms of
the generalized thermodynamics. It follows from the basics
principles of statistical mechanics that one can use the
microcanonical distribution:
\begin{eqnarray}
\rho_E(C):=\rm const \mit \delta(H_0(C)-E) \label{VI}
\end{eqnarray}
instead of the canonical distribution
\begin{eqnarray}
\rho_T=\frac{e^{-\frac{H_0(C)}{T}}}{\rm tr \mit
(e^{-\frac{H_0(C)}{T}})}.
\end{eqnarray}
the entropy corresponding to the distribution has the form:
\begin{eqnarray}
S(E,C)=\rm tr \mit \delta(H_0(C)-E). \label{VII}
\end{eqnarray}
Note that in equation (\ref{VI}) and (\ref{VII}) by symbol
\(\delta(x)\) we mean some regularization of \(\delta\)-function
\(\delta\)-function \(\Delta(x)\) defined as follow:
\begin{eqnarray}
\Delta(x)=0,\; \rm if \mit \;|x|>\frac{\Delta}{2}\nonumber\\
\Delta(x)=\frac{1}{\Delta},\; \rm if \mit \; |x|<\frac{\Delta}{2}
\end{eqnarray}
where \(\Delta\) is some small positive real number (asymptotical
constant as \(V\rightarrow \infty\)).

Let us consider the following symplectic manifold \(\mathfrak{S}_2\)
with the canonically-conjugated variables
\(S_R,\;S_I,\;\Phi_R,\;\Phi_I\) on it defined as follows.

Variables \(S_R,\;S_I\) run the whole real line, but variables
\(\Phi_R\), \(\Phi_I\) are defined modulo the substitutions:
\begin{eqnarray}
\Phi_R\mapsto\Phi_I+2\pi n,\nonumber\\
\Phi_I\mapsto\Phi_I+2\pi n,\nonumber\\
\end{eqnarray}
where \(n\in \mathbb{Z}\). The Poisson brackets of variables
\(S_R,\;S_I,\;\Phi_R,\;\Phi_I\) by definition has the form:
\begin{eqnarray}
(S_R,S_I)=(\Phi_R,\Phi_I)=0,\nonumber\\
(S_R,\Phi_I)=(S_I,\Phi_R)=0,\nonumber\\
(S_R,\Phi_R)=(S_I,\Phi_I)=1.\nonumber\\
\end{eqnarray}
Let \(\mathfrak{S}_1\) be a (non-commutative) phase space of the
dynamical system, described by the Hamiltonian (\ref{Hamiltonian}).
Let \(\mathfrak{S}\) be a direct product of the
(non-commutative)spaces \(\mathfrak{S}_1\) and \(\mathfrak{S}_2\),
\(\mathfrak{S}=\mathfrak{S}_1\times\mathfrak{S}_2\). Now let us
consider the dynamical system on the space \(\mathfrak{S}\) which
described by Hamiltonian \(H_0(S_R+S_I)\). It is evident that the
dynamical variables \(S_R\) and \(S_I\) are the commuting first
integrals with respect to the Hamiltonian \(H_0(S_R+iS_I)\). The
generalized micro-canonical distribution corresponding to the
integral \(H_0(S_R+iS_I)\) and the integrals \(S_R\) and \(S_I\)
could be written as follows:
\begin{eqnarray}
S^\mathfrak{S}(E,C):=\rm tr \mit_{\mathfrak{S}^1} \int
d\Phi_R...dS_I
\delta(H_0(S_R+iS_I)-E)\nonumber\\
\times\delta(S_R-\mathfrak{R}C)\delta(S_I-\mathfrak{I}C)=S(E,C)+\rm
c \mit.
\end{eqnarray}
Here \(S(E,C)\) is defined by equation (\ref{VII}), \(c\) is some
real constant which does not depend on \(E\) and \(C\) and symbol
\(\rm tr \mit_{\mathfrak{S}^1}\) means trace over the Hilbert space
of our model dynamical system described by the Hamiltonian
(\ref{Hamiltonian}). Recall that one of the basics principles of our
generalized thermodynamics state that the observable value of \(C\)
(which may be realized in nature) must corresponds to the maximum of
\(S^\mathfrak{S}(E,C)\) (at fixed energy \(E\)). We have proved that
this condition is equivalent to the equation of compensation
(\ref{Compens}). Therefore, we have formulated the thermodynamics of
our model in terms of generalized thermodynamics. Note also that all
nonzero solutions of equation of compensation form some
one-dimensional manifold. So the entropy \(S^\mathfrak{S}(E,C)\) (at
fixed energy) has one- dimensional plateau.

\section{On some new method of derivation of Boltzmann kinetic equation.}
In the present section we consider Bose gas with weak pair enough
good interaction. For this system we derive usual kinetic Boltzmann
equation in the limit of weak interaction by some new method. We
consider a single mode of Bose gas (corresponding to some momentum
\(k\) as an oscillator interacting with other modes as an oscillator
interacting with thermostat. To study the evolution of this fixed
mode interacting with other modes (considered as thermostat) we will
use the so called statistical perturbation theory [10].

One of the interesting consequence from non-ergoding theorem states
the following. To obtain an irreversible macroscopic evolution from
the reversible microscopic evolution one need to take into account
the behavior of the system at its boundary. In other words it is
impossible to prove that the thermalization takes place in the
system if we neglect by  the role which plays the environment of the
system in this process. The derivation of kinetic equations
presented in this section gives us a new point of view at the role
which plays the environment of the system in the process of its
thermalization.

So let us consider a model dynamical system, described by the
following Hamiltonian,
\begin{eqnarray}
H=H_0+\lambda V, \label{HAM1}
\end{eqnarray}
where \(H_0\) is a free Hamiltonian
\begin{eqnarray}
H_0=\int (\frac{k^2}{2}-\mu)a^+(k)a(k)d^3k \label{HAM2}
\end{eqnarray}
and
\begin{eqnarray}
H_1=\frac{\lambda}{2}\int
d^3p_1d^3p_2d^3q_1d^3q_2v(p_1,p_2|q_1,q_2)\nonumber\\
\times\delta(p_1+p_2-q_1-q_2)a^+(p_1)a^+(p_2)a(q_2)a(q_1)
\label{HAM3}
\end{eqnarray}
Here \(a^+(k),\;a(k)\) are usual Bose creation-annihilation
operators (described in section 2), \(\mu \in \mathbb{R}\) is a
chemical potential \(\lambda \in \mathbb{R}\) is a coupling
constant. \(v(p_1,p_2|q_1,q_2)\) is a test function from the
Schwartz space such that
\begin{eqnarray}
v(p_1,p_2|q_1,q_2)=v^{\ast}(q_1,q_2|p_1,p_2).
\end{eqnarray}
Note that the Hamiltonian which describes Bose gas with weak (enough
good) is a Hamiltonian of such form.

Now let us briefly describe the representation of quasi-particles in
statistical mechanics. We will use this representation below when we
try to establish the form of the density operator for
non-equilibrium state.

Let us consider Bose Gas contained in macroscopic volume \(V\). Let
us formulate the main physical assumption which lays in the
fundament of quasi-particles representation. The classification of
the energetic levels of interacting particle could be described in
the same manner as a classification of energetic levels of the
system of non interacting particles. Therefore the energetic levels
of the system could be described by means of real valued pointwise
positive function \(n(p)\) of momentum \(p\in \mathbb{R}^3\). One
can interpret the density function \(n(p)\) as a density of
quasi-particles with momentum \(p\). In other words \(n(p)d^3p\) is
a number of particles with momenta from infinitesimally small volume
(in momenta space) \(d^3p\) with the center at \(p\). Let \(E[n]\)
be an energy of an energetic level corresponding to the function
\(n(p)\). Let \(\varepsilon(p)\) be a functional derivation of
\(E[n]\) with respect \(n(p)\), i.e.
\begin{eqnarray}
\varepsilon(p,[n])=\frac{\delta E[n]}{\delta n(p)}.
\end{eqnarray}
One can interpret \(\varepsilon(p)\) as a dispersion low of
quasi-particles. We have supposed that the classification of
energetic levels of interacting system is the same as a
classification of energetic levels of non-interacting system.
Therefore the entropy of the system corresponding to the density
function \(n(p)\) has the form:
\begin{eqnarray}
S[n]=\int d^3p[(1+n(p))\ln (1+n(p))-n(p)\ln n(p)].
\end{eqnarray}
Note that the dynamics of (isolated) interacting system differs from
the dynamics of non-interacting systems only in the following/ The
dispersion low for \(\varepsilon(p,[n])\) for interacting system
depends of the density function \(n\).

Therefore for all piece-wise continuous functions \(f(p)\) the
quantities
\begin{eqnarray}
K[f]=\int \hat{n}(p)f(p)d^3p
\end{eqnarray}
are commuting integrals of motion. Here, by definition
\begin{eqnarray}
\hat{n}(p):=\alpha^+(p)\alpha(p)
\end{eqnarray}
and \(\alpha^+(p)\) and \(\alpha(p)\) are formally defined operators
of quasi-particles, satisfying to the Bose-Einstein statistics.

Note that last observation is the main idea of the proof of
non-ergodic theorem. More precisely for the systems which contains
of finite number of particles (quantum fields) wave operator \(W\)
(defined in section 2) maps dynamics of free particles into dynamics
of interacting particles. Therefore it is essential try to find the
construction analogues to the construction of wave operators to
build the quasi-particles representation. But the series of
perturbation theory obtained by such a way contain (secular)
divergences. One of the authors comes to non-ergodic theorem when
tried to renormalize such divergences.

Let \(\{\mathcal{O}_i|i=1,2,3...\}\) is a decomposition of
\(\mathbb{R}^3\) into a set of domains such that \(\forall i\neq j\)
\(\mathcal{O}_i\cap\mathcal{O}_j\subset\partial\mathcal{O}_i\cup\partial\mathcal{O}_j\),
where \(\partial U\) means the boundary of the set \(U\). Suppose
that \(\forall i=1,2,3...N\) and \(\forall p,p' \in \mathcal{O}_i\)
the difference \(p-p'\) is very small. \(\forall i=1,2,3...\) put by
definition
\begin{eqnarray}
K_i=\int \limits_{\mathcal{O}_i}\hat{n}(p)d^3p
\end{eqnarray}
Now let us find the form of distribution function \(n(p)\)
corresponding to some (generally) equilibrium state (in the sense of
generalized microcanonical distribution) which corresponds to fixed
values \(K'_i\) of integrals \({K}_i,\;i=1,2,3,...\).

It is evident that (generally) equilibrium distribution function
\(n(p)\) corresponds to (relative) maximum of the entropy \(S(n)\)
under the following constrains:
\begin{eqnarray}
K_i[n]:=\int d^3p n(p)=K'_i,\nonumber\\
i=1,2,3....
\end{eqnarray}
and
\begin{eqnarray}
E[n]=E.
\end{eqnarray}
According to the method of undefined lagrange multipliers we reduce
this problem to the solution of the following system of equations
\begin{eqnarray}
\frac{\delta}{\delta n(p)}\{-\sum \limits_{i=1}^\infty
G_i[(1+\frac{N_i}{G_i})\ln(1+\frac{N_i}{G_i})-\frac{N_i}{G_i} \ln
\frac{N_i}{G_i}]\nonumber\\
-\sum \limits_{i=1}^N \mu_i N_i+E[n]\}=0,
\end{eqnarray}
where
\begin{eqnarray}
G_i:=\int \limits_{\mathcal{O}_i} d^3p
\end{eqnarray}
and
\begin{eqnarray}
N_i=\int \limits_{\mathcal{O}_i} n(p) d^3p, \nonumber\\
i=1,2,3...
\end{eqnarray}
This system of equation could be immediately solved and one finds
\begin{eqnarray}
n(p)=\frac{1}{e^{\frac{\varepsilon(p,[n])-\mu(p)}{T}}-1},
\end{eqnarray}
where
\begin{eqnarray}
\mu(p)=\mu_i \in \mathbb{R},\; \rm if \mit p\; \in \mathcal{O}_i
\end{eqnarray}
and \(T\) some positive number

In other words if one moves from generalized microcanonical
distribution to the generalized microcanonical distribution this is
equivalent to the following. The chemical potential \(\mu\) becomes
to the function of momenta \(p\).

Note that in spite of volume \(V\) is macroscopic it is finite.
Therefore the set of all possible momenta is discrete
\(\{p_i|i=1,2,3...\}\). Let by definition \(n_i\) be a number of
particles with momentum \(p_i\).

\(\forall i=1,2,3,...\) the cells \(\mathcal{O}_i\) corresponds to
the approximately equal frequencies. Therefore each two states
corresponding to the same numbers \(N_i,\;i=1,2,3,....\) corresponds
to approximately the same energy. Therefore all the states
corresponding to the same numbers \(N_i\;i=1,2,3,...\) have
approximately the same probabilities.

One can easily find the form of the distribution function
\(\rho_i(n_i)\) by using the method of the most probable
distribution. Omitting the well-known reasoning one finds
\begin{eqnarray}
\rho_i(n_i)=\frac{e^{-\frac{\varepsilon(p_i,[n])n_i-\mu_i
n_i}{T}}}{Z_i},
\end{eqnarray}
where \(Z_i\) is a real positive number such that
\begin{eqnarray}
\sum \limits_{n=0}^{\infty}\rho_i(n)=1.
\end{eqnarray}

Therefore the (generalized) equilibrium state of quasiparticles
could be described by means the Gauss state on the algebra of
canonical commutative relations generated by \(\alpha(p)\),
\(\alpha^+(p)\) of the form
\begin{eqnarray}
\langle\alpha(p)\alpha(p')\rangle=\langle\alpha^+(p)\alpha^+(p')\rangle=0,\nonumber\\
\langle\alpha^+(p)\alpha(p')\rangle=\delta(p-p')(1+n(p)),\nonumber\\
n(p)=\delta(p-p')\frac{1}{e^{-\frac{\varepsilon(p,[n])-\mu(p)}{T}}-1},
\end{eqnarray}
\begin{eqnarray}
\mu(p)=\mu_i,\;\rm if \mit p \in \mathcal{O}_i.
\end{eqnarray}

So, let us consider a single mode of momenta \(k_i\) interacting
with other modes of Bose gas as with a thermostat. To analyze this
system we will use so-called statistical perturbation theory [10].
Bogoliubov [10] has considered the following system. He has
considered the Hamiltonian system of \(n\) degrees of freedom
described by the Hamiltonian \(H(p,q)\) in the field of external
force \(\varepsilon f(t)\), \(\varepsilon \in \mathbb{R}\). So,
total Hamiltonian of the system has the form
\begin{eqnarray}
\Gamma=H+\varepsilon f(t)P,
\end{eqnarray}
where \(P(p,q)\) is some function of canonically-conjugated of
momenta and coordinates. We also suppose that external force
\(f(t)\) is the function of time \(t\) of the form:
\begin{eqnarray}
f(t)=\sum \limits_\nu a_\nu \cos (\nu t+\varphi_\nu).
\end{eqnarray}
Here the phases \(\varphi_\nu\) are independent random quantities
uniformly distributed on circle. We also suppose that the spectrum
of frequencies is almost continuous so we can the sums of the form
\begin{eqnarray}
\sum \limits_\nu F(\nu) a_\nu^2
\end{eqnarray}
for continuous functions \(F(\nu)\) replace by the integral
\begin{eqnarray}
\int \limits_0^{+\infty}F(\nu)I(\nu)d\nu.
\end{eqnarray}

Let \(D_t\) denote the probability density of momenta and
coordinates at fixed time moment \(t\) under the condition that
phases \(\varphi_\nu\) take some defined values. The probability
density of momenta and coordinates in usual sense (for arbitrary
values of phases \(\varphi_\nu\)) could be obtained from \(D_t\) by
means of averaging over the phases \(\varphi_\nu\). In other words
\begin{eqnarray}
D_0=\rho_0.
\end{eqnarray}
The evolution of \(D_t\) in times could be described by means well
known Liouville equation
\begin{eqnarray}
\frac{\partial D_t}{\partial t}=(H,D_t)+\varepsilon f(t)(P,D_t),
\end{eqnarray}
where \((A,B)\) denotes usual Poisson Brackets defined by the
formula
\begin{eqnarray}
(A,B)=\sum \limits_{i=1}^n (\frac{\partial A}{\partial
q_i}\frac{\partial B}{\partial p_i}-\frac{\partial A}{\partial
p_i}\frac{\partial B}{\partial q_i}).
\end{eqnarray}
Let us introduce the following one-parameter group of operators
\(T_t\) acting on dynamical variables according to the formula
\begin{eqnarray}
T_t F(p,q)=F(p_t,q_t),
\end{eqnarray}
where \((p_t,q_t)\) are the solution of canonical Hamiltonian
equations
\begin{eqnarray}
\frac{dp_t}{dt}=-\frac{\partial H(p_t,q_t)}{\partial
q_t},\nonumber\\
\frac{dq_t}{dt}=\frac{\partial H(p_t,q_t)}{\partial p_t},
\end{eqnarray}
corresponding to the following initial conditions
\begin{eqnarray}
p_0=p,\nonumber\\
p_0=p.
\end{eqnarray}
In these assumptions and denotations in the limit of small coupling
constant \(\varepsilon\) it is derived [Bogoliubov, 1945] the
following equation on \(\rho_t\)
\begin{eqnarray}
\frac{\partial\rho_t}{\partial t}=(H,\rho_t)+\varepsilon^2 \int
\limits_0^t \Delta(t-\tau)(P,(T_{t-\tau}(P,\rho_\tau)))d\tau,
\label{main}
\end{eqnarray}
where
\begin{eqnarray}
\Delta(\tau)=\frac{1}{2} \int \limits_0^{+\infty}I(\nu) \cos (\nu
\tau)d\tau.
\end{eqnarray}
The equation (\ref{main}) could be essentially generalized to the
case of quantum mechanics. It is enough to replace the Poisson
brackets by the commutator.

Note that the random force \(f(t)\) here is an example of random
processes. Recall that the random process is the set \((T,f_t)\),
where \(T=(S,\Sigma,P)\) is a Kolmogorov triple (\(S\) is some set,
\(\Sigma\) is a \(\sigma\)-algebra, \(P\) is a probability measure
on \(S\) defined on sets from \(\Sigma\)). \(f_t\) is a map which to
an arbitrary \(t \in\mathbb{R}\) assigns a random quantity i.e. a
measurable (with respect to \(\Sigma\)) function on \(S\). Note that
the equation (\ref{main}) could be rewritten as follows
\begin{eqnarray}
\frac{\partial \rho_t}{\partial t}=(H,\rho_t)+\varepsilon^2 \int
\limits_0^t \overline{f(t)t(\tau)}(P,(T_{t-\tau}(P,\rho_t)))d\tau,
\end{eqnarray}
where \(\overline{\xi}\) means the mathematical expectation of
\(\xi\) i.e.
\begin{eqnarray}
\bar{\xi}:=\int \limits_S \xi(x)dP(x).
\end{eqnarray}
Note that how it follows from the derivation of equation
(\ref{main}) [Bogoliubov, 194] it is not necessary that \(f(t)\) has
the form described previously. It is enough that \(f(t)\) has the
form
\begin{eqnarray}
f(t)=\sum \limits_\nu \frac{1}{2}(f_\nu e^{i\nu t}+f^{\ast}_\nu
e^{-i\nu t}),
\end{eqnarray}
where \(f_\nu\) are random quantities such that \(\forall \mu,
\nu\),
\begin{eqnarray}
\overline{f_\mu}=\overline{f_\mu^\ast}=0,
\end{eqnarray}
\begin{eqnarray}
\overline{f_\mu f_\nu}=\overline{f_\mu^\ast f^\ast_\nu}=0,
\end{eqnarray}
and
\begin{eqnarray}
\overline{f_\nu f_\mu}=\delta_{\mu \nu} I_\nu.
\end{eqnarray}
The spectrum of frequencies  \(\{\nu\}\) is supposed almost
continuous such that one can replace the sums
\begin{eqnarray}
\sum \limits_\nu I_\nu F(\nu)
\end{eqnarray}
for continuous functions \(F(\nu)\) by the integral
\begin{eqnarray}
\int \limits_0^{+\infty} I(\nu)F(\nu) d\nu.
\end{eqnarray}
Now note that the generalization of a Kolmogorov triple to
noncommutative (quantum) case is a pair consisting of an unital
\(\star\)-algebra \(\mathcal{A}\) and a state
\(\langle\cdot\rangle\) on it. By definition state is a normalized
positively defined linear functional on the algebra. Normalization
condition means that \(\langle\mathbf{1}\rangle=1\), where
\(\mathbf{1}\) is a unite of \(\mathcal{A}\). Positively
definiteness means that \(\forall a \in \mathcal{A}\) \(\langle a
a^{\star}\rangle\geq 0\).

The algebra \(\mathcal{A}\) plays the role of the space of
measurable functions on (nonexisting) noncommutative space. The
state \(\langle\cdot\rangle\) plays the role of integration on this
state. The involution \(\star\) plays the role of pointwise complex
conjugation.

By definition the noncommutative random process is a map which to an
arbitrary \(t \in \mathbb{R}\) assigns an element \(a_t\) of
\(\mathcal{A}\). We say that noncommutative random process \(\{a_t|t
\in \mathbb{R}\}\) is  a real-valued process if \(\forall t \in
\mathbb{R}\) \(a_t=a^{\star}_t\).

The equation (\ref{main}) could be generalized to the case of
noncommutative (quantum) systems. More precisely let total
Hamiltonian of the system has the form
\begin{eqnarray}\
H=H_0+\varepsilon(f(t)P+f^+(t)P^+),
\end{eqnarray}
where  \(f(t)\) is a noncommutative random process of the form
\begin{eqnarray}
f(t)=\sum \limits_\nu f_\nu e^{i\nu t},
\end{eqnarray}
\(\forall \mu\),
\begin{eqnarray}
\langle f_\mu\rangle=\langle f_\mu^\star\rangle=0,
\end{eqnarray}
and \(\forall \mu,\nu\)
\begin{eqnarray}
\langle f_\mu f_\nu\rangle=\langle f_\mu^\star f_\nu^\star\rangle=0,
\label{PAIRS1}
\end{eqnarray}
\begin{eqnarray}
\langle f_\mu^\star f_\nu  \rangle=I_\nu\delta_{\mu \nu},\nonumber\\
\langle f_\mu f_\nu^\star  \rangle=J_\nu\delta_{\mu
\nu}.\label{PAIRS2}
\end{eqnarray}
The spectrum of external force \(f(t)\) is assumed almost continuous
such that we can replace the sums
\begin{eqnarray}
\sum \limits_\nu F(\nu)I_v,\nonumber\\
\sum \limits_\nu F(\nu)J_\nu
\end{eqnarray}
by the integrals
\begin{eqnarray}
\int \limits_0^{+\infty} F(\nu)I(\nu)d\nu,\nonumber\\
\int \limits_0^{+\infty} F(\nu)J(\nu)d\nu
\end{eqnarray}
respectively. In this case, instead of (\ref{main}) we will have
\begin{eqnarray}
\frac{\partial\rho_t}{\partial t}=[H,\rho_t]\nonumber\\
+\varepsilon^2 \int
\limits_0^t\langle[f^\star(t)P^\star,f(\tau)T_{\tau-t}([P,\rho_t])]\rangle\nonumber\\
+\varepsilon^2 \int
\limits_0^t\langle[f(t)P,f^\star(\tau)T_{\tau-t}([P^\star,\rho_t])]\rangle
\label{Equation} .
\end{eqnarray}

We suppose that our system is contained in some cube of volume \(V\)
with edges parallel to the coordinate axis. We suppose that the
length od the edges of this cube is equal to \(L\). Therefore
\(V=L^3\). Put by definition
\begin{eqnarray}
\mathfrak{K}=\{k=(k_x,k_y,k_z)|\frac{k_x}{2\pi L},\frac{k_y}{2\pi
L},\frac{k_z}{2\pi L}\in \mathbb{Z}\}.
\end{eqnarray}
We need to use the discrete momentum representation for finite
volume instead of continuous momentum representation for infinite
volume. Instead of the Hamiltonian (\ref{HAM1},\ref{HAM2},
\ref{HAM3}) we need to use the following Hamiltonian in finite
volume:
\begin{eqnarray}
H^V=H_0^V+H_1^V,
\end{eqnarray}
where
\begin{eqnarray}
H_0^V=\sum \limits_{k \in \mathfrak{K}} \omega(k)a^+(k)a(k),
\end{eqnarray}
and
\begin{eqnarray}
\omega(k)=\frac{k^2}{2}.
\end{eqnarray}
\begin{eqnarray}
H_1^V=\frac{\lambda(2\pi)^3}{2V} \sum \limits_{p_1,p_2;q_1,q_2 \in
\mathfrak{K}} \Delta(p_1+p_2-q_1-q_2)v(p_1,p_2|q_1,q_2)\nonumber\\
\times a^+_{p_1}a^+_{p_2}a_{q_2}a_{q_1},
\end{eqnarray}
where
\begin{eqnarray}
\Delta(p)=1,\;\rm if \mit\; p=0\nonumber\\
\Delta(p)=0,\;\rm if \mit\; p\neq 0,
\end{eqnarray}
and \(a_p^+\) and \(a_p\) are creation and annihilation operators in
discrete momentum representation.

Now let us consider a single mode corresponding to some fixed
momentum \(k \in \mathfrak{K}\). Let us study how the state of this
mode changes while time changes from \(t\) to \(t+dt\) for
infinitely small \(dt\). To study how the density matrix \(\rho_t\)
of this mode changes at time interval \([t,t+dt]\) one can suppose
that the thermostat evolves at this interval according to free
dynamics. So one can suppose that the mode corresponding to the
momentum \(k\) affected by external (noncommutative) force defined
by this thermostat. In other words we must take
\begin{eqnarray}
\Gamma=H+\lambda (f(t)P+f^{\star}(t)P^+),
\end{eqnarray}
where, by definition
\begin{eqnarray}
H=\omega(k)a^+(k)a(k), \nonumber\\
P=a_k.
\end{eqnarray}
External forces \(f(t)\) are the elements of the algebra
\(\mathcal{A}\) generated by operators \(a^+_p,\;a_p\), \(p\neq k\).
External force \(f(t)\) has the form
\begin{eqnarray}
f(t)=\frac{(2\pi)^3}{V} \sum \limits_{p_1,p_2,q_1 \in
\mathfrak{K}\setminus\{k\}}
\Delta(p_1+p_2-q_1-k)v(p_1,p_2|q_1,q_2)\nonumber\\
\times a^+_{p_1}a^+_{p_2}a_{q_1}
e^{i(\omega(p_1)+\omega(p_2)-\omega(q_1))t}
\end{eqnarray}
To obtain the differential equation for the density matrix for our
fixed mode we will average over the state \(\langle\cdot\rangle\) on
the algebra \(\mathcal{A}\) which is a Gauss state and is defined by
the following its pair correlator
\begin{eqnarray}
\langle a_pa_{p'}\rangle=\langle a^+_pa^+_{p'}\rangle=0,\nonumber\\
\langle a^+_p a_{p'}\rangle=\Delta(p-p')n(p).
\end{eqnarray}
Note that external force \(f(t)\) has the form
\begin{eqnarray}
f(t)=\sum \limits_\nu f_\nu e^{i\nu t},
\end{eqnarray}
where
\begin{eqnarray}
f_\nu=\frac{(2\pi)^3}{V} \sum \limits_{p_1,p_2,q_1 \in
\mathfrak{K}\setminus\{k\}}
\Delta(p_1+p_2-q_1-k)\nonumber\\
\times\Delta(\omega(p_1)+\omega(p_2)-\omega(q_1)-\nu)
v(p_1,p_2|q_1,q_2)a^+_{p_1}a^+_{p_2}a_{q_2}.
\end{eqnarray}
Note that the averaged pair products of \(f_\nu,\;f_\nu^\star\)
satisfy to the equations (\ref{PAIRS1},\ref{PAIRS2}). Let \(\rho_t\)
be a density matrix corresponding to our fixed single mode with
momentum \(k\). Put by definition \(\langle\cdot\rangle=\rm tr \mit
(\cdot \rho_t)\).

Suppose that \(\langle\cdot\rangle\) be a Gauss state on the algebra
\(\mathcal{A}\). Equation (\ref{Equation}) implies that \(\forall
t>0\) the state \(\langle\cdot\rangle_{t+dt}\) is a Gauss state if
the state \(\langle\cdot\rangle_t\) is a Gauss state for infinitely
small positive \(dt\). Therefore if \(\langle\cdot\rangle_t\) is a
gauss state for all positive \(t\) if \(\langle\cdot\rangle_0\) is a
Gauss state (in the limit of weak pair interaction). Let
\(\hat{n}_p=a^+_pa_p\) be an operator of the particle number
corresponding to the momentum \(p\). Put by definition
\(n_t(p)=\langle \hat{n}_p\rangle_t\). Equation (\ref{Equation})
implies that (in the limit of weak pair interaction)
\begin{eqnarray}
\frac{d}{dt}n_t(k)=-\lambda^2 \int
\limits_0^t\langle\langle[a_k^+f^\star(t),a_kf(\tau)]\rangle_t\rangle
e^{i\omega(k)(t-\tau)}d \tau \nonumber\\
+\lambda^2 \int
\limits_0^t\langle\langle[a_kf(t),a^+_kf^\star(\tau)]\rangle_t\rangle
e^{- i\omega(k)(t-\tau)}d \tau\nonumber\\
=-\lambda^2\int_0^t\{n_t(k)\langle
f^\star(t)f(\tau)\rangle-(1+n_t(k))\langle
f(\tau)f^\star(t)\rangle\}e^{i\omega(k)(t-\tau)}d \tau \nonumber\\
+\lambda^2\int_0^t\{(1+n_t(k))\langle
f(t)f^\star(\tau)\rangle-n_t(k)\langle
f^\star(\tau)f(t)\rangle\}e^{-i\omega(k)(t-\tau)}d \tau \nonumber\\
=-\lambda^2 n_k(t) \int \limits_{-t}^t\langle
f^\star(0)f(\tau)\rangle e^{-i\omega(k)\tau}d\tau \nonumber\\
+\lambda^2 (1+n_k(t)) \int \limits_{-t}^t\langle
f(0)f^\star(\tau)\rangle e^{i\omega(k)\tau}d\tau.
\end{eqnarray}
Therefore
\begin{eqnarray}
\frac{d}{dt}n_t(k)\nonumber\\
=-\frac{\lambda^2(2\pi)^6}{V^2} \sum \limits_{p_1,p_2,q_1
\in\mathfrak{K}\backslash
\{k\}}|v(p_1,p_2|q_1,k)|n_t(q_1)n_t(k)(1+n_t(p_1))(1+n_t(p_2))\nonumber\\
\times \Delta(p_1+p_2-q_1-k) \int \limits_{-t}^{t}
e^{i(\omega(p_1)+\omega(p_2)-\omega(q_1)-\omega(q_2))\tau}d \tau
\nonumber\\
+\frac{\lambda^2(2\pi)^6}{V^2} \sum \limits_{p_1,p_2,q_1
\in\mathfrak{K}\backslash
\{k\}}|v(p_1,p_2|q_1,k)|(1+n_t(q_1))(1+n_t(k))n_t(p_1)n_t(p_2)\nonumber\\
\times \Delta(p_1+p_2-q_1-k) \int \limits_{-t}^{t}
e^{-i(\omega(p_1)+\omega(p_2)-\omega(q_1)-\omega(q_2))\tau}d \tau
\end{eqnarray}
But volume \(V\) is macroscopic. So we can replace the sums over
\(\mathfrak{K}\) by integrals. In result
\begin{eqnarray}
\frac{d}{dt}n_t(k)=(2\pi)\lambda^2 \int
d^2p_1d^3p_2d^3q_1\delta(p_1+p_2-q_1-q_2)\nonumber\\
 \delta(\omega(p_1)+\omega(p_2)-\omega(q_1)-\omega(k))
\times\{(1+n_t(q_1))(1+n_t(k))n_t(p_1)n_t(p_2)\nonumber\\
-n_t(q_1)n_t(k)(1+n_t(p_1))(1+n_t(p_2))\}.
\end{eqnarray}
So we have obtain the usual kinetic equation for Bose gas in the
limit of weak interaction.

\section{Integrals for non-linear partial differential equations.}
\textbf{Solitary perturbations from generalized thermodynamics.} Let
us consider some system of statistical mechanics such that there
second-order phase transition takes place at some temperature
\(T_{cr}\). Suppose that this transition concerning with some
symmetry breaking. As an example of such system one can take some
superfluid or superconductive system. Let \(\langle\cdot\rangle\) be
(generally) equilibrium translation-invariant stationary state
corresponding to the temperature \(T<T_{cr}\). Suppose that
\(\langle\vec{P}\rangle=0\), where \(\langle\vec{P}\rangle\) is a
momentum operator. Let \(\mathcal{K}\) be some inertial reference
system which moves with velocity \(v\). Let \(U\subset\mathbb{R}^3\)
be some domains of microscopical sizes tightly connected with
reference system \(\mathcal{K}\). Domain \(U\) by definition is
macroscopic. Therefore, according yo the non-ergodic theorem [1]
there exist integrals \(K_1,...,K_N\) in involution for the part of
our system contained in \(U\). According to our generalized
thermodynamics, the distribution function for the part of our
system, contained in \(U\) has the form:
\begin{eqnarray}
\rho(x)=\rm const \mit \delta(H(x)-E) \prod
\limits_{i=1}^N\delta(K'_i(x)-K'_i).
\end{eqnarray}
Now let us show, that the entropy of the system contained in \(U\):
\begin{eqnarray}
S(E,K'_1,...,K'_n)=\ln \int d\Gamma_x \delta(H(x)-E) \prod
\limits_{i=1}^N\delta(H(x)-E) \prod
\limits_{i=1}^N\delta(K_i(x)-K'_i)
\end{eqnarray}
achieve a maximum at some \(D>0\) dimensional manifold \(M\) as a
function of integrals \(K_1,...,K_N\) at fixed energy \(E\). Indeed
suppose that the entropy achieve a maximum (at fixed energy) at
isolated point. In this case the (generalized) microcanonical
distribution gives us the same results as standard microcanonical
distribution
\begin{eqnarray}
\rho(x)=\rm const \mit \delta(H(x)-E).
\end{eqnarray}
But for this distribution all the symmetries of the Hamiltonian
\(H(x)\) are preserved. Therefore the entropy \(S(E,K'_1,...,K'_N)\)
achieve a maximum at some \(M>0\) dimensional manifold as function
of \(K'_1,...,K'_N\).

Therefore we can chose the observable values \(K'_1,...,K'_N\) of
the integrals \(K_1,...,K_N\) such that (the intensive) properties
of the system will be qualitatively different from the analogues
properties of other part of the system (outside of \(U\)) and values
\(K'_1,...,K'_N\) corresponds to the maximum of the entropy.
According to our generalized thermodynamics this state is stable.
For example, for superfluid gases, the the phase of condensate wave
function inside \(U\) may differ from the phase of condensate wave
function outside of \(U\). If we return into initial (static)
reference system we obtain a non-trivial perturbation of the state
\(\langle\cdot\rangle\) which moves with constant velocity without
dissipation. Such perturbation we will call thermodynamical
solitons.

\textbf{An example of thermodynamical soliton. The Abrikosov
Vortices.} As an example of thermodynamical soliton we consider the
situation which arise in the superfluidity theory if one studies the
condensate wave function [11].

Let \(\hat{\Psi}(x)\) and \(\Psi(x)\) be secondary quantized wave
function and its hermitian-conjugated. Let \(\hat{\Xi}\) and
\(\hat{\Xi}^+(x)\) be the parts of the function \(\hat{\Psi}(x)\)
and \(\hat{\Psi}^+(x)\) which change by one number of condensate
particles and preserve others quantum numbers. In other words, by
definition:
\begin{eqnarray}
\hat{\Xi}(x)|N+1,m\rangle=\Sigma(x)|N,m\rangle,\nonumber\\
\hat{\Xi}^+(x)|N,m\rangle=\Sigma^{\ast}(x)|N+1,m\rangle
\end{eqnarray}
Here
\begin{eqnarray}
\Xi(x):=\langle N,m|\Xi(x)|N+1,m\rangle,
\end{eqnarray}
and \(\Xi^{\ast}(x)\) is a complex-conjugated to \(\Xi(x)\)
function. \(N\) is a number of condensate particle \(m\) is a number
of other quantum numbers of the system. Let us represent
\(\hat{\Psi}(x)\) and \(\Psi(x)\) as follows:
\begin{eqnarray}
\hat{\Psi}(x)=\Xi(x)+\hat{\Psi}'(x), \nonumber\\
\hat{\Psi}^+(x)=\Xi^+(x)+\hat{\Psi}'^+(x).
\end{eqnarray}
In thermodynamical limit \(N\rightarrow \infty\) the difference
between the states \(|N,m\rangle\) and \(|N+1,m\rangle\) disappears
at all. Therefore the operators \(\hat{\Xi}(x)\) and
\(\hat{\Xi}^+(x)\) becomes the operators which commutes to each
other and commutes with \(\hat{\Psi}'(x)\) and \(\hat{\Psi}'^+(x)\).
In other words \(\hat{\Xi}(x)\) and \(\hat{\Xi}^+(x)\) becomes the
classical variables. Now let us write the (partial) differential
equation on condensate wave function \(\Xi(x)\). Let us consider a
weakly non-ideal Bose gas at absolute zero of temperature. Almost
all particles in such a gas are in condensate state. In terms of
secondary-quantized wave function this means that the
over-condensate part \(\hat{\Psi}'(x)\) is small with respect to the
condensate wave function \(\hat{\Xi}(x)\). If we neglect by small
over-condensate part \(\hat{\Psi}'(x)\) the condensate wave function
will satisfy to the same Schrodinger equation as the equation which
take place for the operator \(\hat{\Psi}(x)\). I.e.
\begin{eqnarray}
i\frac{\partial}{\partial
t}\Xi(x,t)=(-\frac{1}{2m}\nabla^2+\mu)\Xi(x,t)\nonumber\\
+\Xi(x,t)\int |\Xi(x',t'|^2U(x-x')d^3x'. \label{Pita}
\end{eqnarray}
Here \(U(x)\) is an interaction potential \(\mu=nU_0\), \(n\) is a
density of particles in gas and \(U_0:=\int U(x) d^3x\). This
equation is called the Pitaevskii equation. We will show now that
the Pitaevskii equation will precise in the limit when the
interaction constant tends to zero. Let us suppose that the
interaction potential \(U(x)\) depends on small positive parameter
\(\lambda\) as follows
\begin{eqnarray}
U(x|\lambda):=\lambda\lambda^{3/2}U(x\sqrt{\lambda}).
\end{eqnarray}
For example if \(U(x)=\delta (x)\) then
\(U(x,\lambda)=\lambda\delta(x)\). Let us introduce new condensate
wave function:
\begin{eqnarray}
\Sigma'(x,t)=\Sigma({x}{\sqrt{\lambda}},t{\lambda}).
\end{eqnarray}
In the limit \(\lambda\rightarrow 0\), \(\Sigma'(x,t)\) will satisfy
precisely to the Pitaevskii equation (\ref{Pita}). The fact that the
Pitaevskii equation will be satisfied precisely will follows from
the fact that all particles will be in condensate in limit
\(\lambda\rightarrow 0\) and the over-condensate wave function will
be equal to zero in the limit \(\lambda\rightarrow 0\).

\textbf{The (partial) differential equation for the order parameter
of the system has a lot of non trivial first integrals.} In this
subsection we consider only super fluid systems. These systems are
described by the condensate wave function \(\Xi(x)\) defined on
\(\mathbb{R}^3\). The condensate wave function play the role of
order parameter for the system. In the limit \(\lambda\rightarrow
0\) the condensate wave function \(\Xi(x)\) and its complex
conjugated \(\Xi^{\ast}(x)\) are the canonically-conjugated
coordinates on some (infinitely-dimensional) phase space. The
Poisson brackets of \(\Xi(x)\) and \(\Xi^{\ast}(x)\) has the form:
\begin{eqnarray}
(\Xi(x),\Xi^{\ast}(x'))=\delta(x-x'). \label{Poisson}
\end{eqnarray}
The evolution of the condensate wave function \(\Xi(x)\) is a
Hamiltonian evolution with respect to the Poisson brackets
(\ref{Poisson}). The corresponding Hamiltonian (for an arbitrary
temperature) is \(F(\Xi(x),\Xi^+(x))|T)\), where
\(F(\Xi(x),\Xi^+(x))|T)\) is a free (Gibbsian) energy of the system
under the condition of fixed condensate wave function \(\Xi(x)\).
Now we will introduce the assumption of so called asymptotically
completeness and prove that the equation for order parameter has a
lot of first integrals in involution under this assumption.

\textbf{Asymptotical completeness.} Let \(\Xi(x,t)\) be a solution
of the Pitaevskii equation (\ref{Pita}) which is localized (in some
essential non rigorous sense) in some  finite domain of the space at
\(t=0\). We ssaume that this solution splits as \(t\rightarrow
\infty\) into the set of separated thermodynamical solitons
described in the beginning of this section.

So let us suppose that the asymptotical completeness takes place. In
this case first integrals for the system, defined by the Hamiltonian
\(F(\Xi(x),\Xi^+(x))|T)\) (for the temperature \(T=0\)) could be
described as follows.

Let \(L\) be a cubic lattice in \(\mathbb{R}^3\). Let
\(\{C_i|\;i=1,2,3,...\}\) be the set of all elementary cubes of
\(L\). We suppose that the sizes of cubes \(C_i\) are macroscopic.
We will also denote by \(C_i\) (\(\forall i=1,2,3,...)\) the part of
our system contained in the cube \(C_i\). \(\forall i=1,2,...\)
system \(C_i\) is macroscopic. So Let \(\{K_i^j\}\), \(j=N\) be some
set of its first integrals in involution. We have shown [2] that
number \(N\) could be chosen as large as needed if each system
\(C_i\) is enough large. Let \(\forall \vec{a} \in \mathbb{R}^3\)
\(T_a:\mathbb{R}^3\rightarrow\mathbb{R}^3\) is a map defined as
follows:
\begin{eqnarray}
T_a:x\mapsto x+\vec{a}.
\end{eqnarray}
\(\forall \vec{a} \in \mathbb{R}^3\) put by definition
\(C(\vec{a})=T_{\vec{a}} C\).

Let \(\vec{a}\) be an arbitrary vector from \(\mathbb{R}^3\). Denote
by \(K_i^j(a)\), \(j=1,2,3,...\) the integrals of the system
\(C_i(\vec{a})\) obtained from \(K_i^j\) by translation on vector
\(\vec{a}\) in obvious sense.

Let \(\forall \vec{a} \in \mathbb{R}^3\)
\(\mathcal{K}_i^j(\vec{a})=\langle
{K}_i^j(\vec{a})\rangle_{\Xi,\Xi^{\ast},T=0}\). Here \(\langle
\cdot\rangle_{\Xi,\Xi^{\ast},T}\) by definition is an averaging by
relative Gibbs distribution corresponding to fixed \(\Xi(x)\) and
\(\Xi^{\ast}\) and the temperature \(T\).

Now let \(\Xi(x)\) be some localized in the space function. Let
\(\Xi(x,t)\) be a solution of Pitaevskii equation such that
\(\Xi(x,0)=\Xi(x)\). Recall that Pitaevskii equations are
Hamiltonian equations with respect to Hamiltonian
\(F(\Xi(x),\Xi^+(x))|T)\). According to the asymptotic completeness
assumption if chose enough large time \(t\) we can assume that
\(\Xi(x,t)\) is a set of space separated thermodynamical solitons
such that for each \(i=1,2,3...\) each cube \(C_i\) has nontrivial
intersection at most with one of such solitons. \(\forall
j=1,2,...N\) put by definition:
\begin{eqnarray}
\mathcal{K}_j(\Xi,\Xi^*)=\int \limits_{C_1} d^3\vec{a} \sum
\limits_{i=1}^{\infty}\mathcal{K}_i^j(\vec{a}(\Xi(x,t),\Xi^{\ast}(x,t)).
\end{eqnarray}
It is evident that this definition do not depend on \(t\) (if \(t\)
is enough large) and \(\mathcal{K}_j(\Xi,\Xi^*)\) are integrals of
motion.

\textbf{Theorem.} Integrals \(\mathcal{K}_j(\Xi,\Xi^*)\),
\(j=1,2,...,N\) are in involution i.e. commutes to each other.

\textbf{Proof.} Suppose that the time \(t\) is chosen as above. We
have:
\begin{eqnarray}
(\mathcal{K}_m(\Xi,\Xi^{\ast}),\mathcal{K}_n(\Xi,\Xi^{\ast}))\nonumber\\
=\int \int_{C_1\times C_1}d^3\vec{a}_1d^3\vec{a}_2 \sum
\limits_{i,j=1,2,...}(\mathcal{K}_m^i(\Xi(t),\Xi^{\ast}(t)),\mathcal{K}_n^j(\Xi(t),\Xi^{\ast}(t)))\nonumber\\
=\int \int_{C_1\times C_1}d^3\vec{a}_1d^3\vec{a}_2\int
\int_{\mathbb{R}^3\times\mathbb{R}^3}d^3x_1d^3x_2 \nonumber\\
\times\sum
\limits_{i,j=1,2,...}\frac{\delta\mathcal{K}_m^i(\Xi(t),\Xi^{\ast}(t))}{\delta\Xi(x_1)},\frac{\delta\mathcal{K}_n^j(\Xi(t),\Xi^\ast(t))}{\delta\Xi^{\ast}(x_2)}
\times (\Xi(x_1),\Xi^{\ast}(x_2))\nonumber\\
+\int \int_{C_1\times C_1}d^3\vec{a}_1d^3\vec{a}_2\int
\int_{\mathbb{R}^3\times\mathbb{R}^3}d^3x_1d^3x_2 \nonumber\\
\times\sum
\limits_{i,j=1,2,...}\frac{\delta\mathcal{K}_m^i(\Xi(t),\Xi^{\ast}(t))}{\delta\Xi^{\ast}(x_1)},\frac{\delta\mathcal{K}_n^j(\Xi(
\Xi(x_2)}{\delta\Xi(x_2)} \times(\Xi^{\ast}(x_1),\Xi(x_2)).
\end{eqnarray}
Therefore to prove that the Poisson Bracket
\((\mathcal{K}_m(\Xi,\Xi^{\ast}),\mathcal{K}_n(\Xi,\Xi^{\ast}))\) is
equal to zero it is enough to prove that the following equalities
holds:
\begin{eqnarray}
\frac{\delta\mathcal{K}_m^i(\Xi,\Xi)}{\delta\Xi(x)}=\frac{\delta\mathcal{K}_m^i(\Xi,\Xi^{\ast})}{\delta\Xi^{\ast}(x)}=0
\end{eqnarray}
for all \(i=1,2,3...\), \(j=1,...,N\), \(\vec{a}, x \in
\mathbb{R}^3\). Let us prove second of these equalities. Second of
these equalities could be proved by analogy.

Fix a cube \(C_i(\vec{a})\). Suppose that the soliton (at most one)
which has nontrivial intersection with \(C_i(\vec{a})\) has zero
velocity. In the opposite case we can chose an inertial system such
that the soliton has zero velocity in this new system. Then we just
perform all reasoning below in this moving system.

The (non-commutative) phase space of the system \(C_i(\vec{a})\)
could be represented as a direct product of two non-commutative
phase spaces \(\mathfrak{S}_1\times\mathfrak{S}_2\) where the phase
space \(\mathfrak{S}_1\) is a classical space and \(\Xi(x)\) and
\(\Xi^*(x)\) are the set of canonically-conjugated coordinates on
it. And the space \(\mathfrak{S}_2\) roughly speaking corresponds to
over-condansate particles. Let \(\mathcal{H}\) be a Hilbert space
corresponding \(\mathfrak{S}_1\). Hamiltonian \(H\) and the
integrals \(K_j\) of the system \(C_i(\vec{a})\) are the functionals
of \(\Xi(x)\) and \(\Xi^*(x)\) which take values in the space of
operators acting in \(\mathcal{H}\). We have
\begin{eqnarray}
\mathcal{K}_m^i(\Xi,\Xi^{\ast})=\frac{{\rm tr(
\mit}K_m(\Xi,\Xi^{\ast}){\rm exp
\mit}(-\frac{H(\Xi,\Xi^{\ast})}{T}))}{\rm tr \mit (\rm exp
\mit(-\frac{H(\Xi,\Xi^{\ast})}{T})) }.
\end{eqnarray}
Here trace is taken over the space \(\mathcal{H}_i(\vec{a})\). We
have
\begin{eqnarray}
\frac{\delta\mathcal{K}_m^i(\Xi,\Xi)}{\delta\Xi(x)}=\frac{{\rm tr
\mit}\{\frac{\delta K_m(\Xi,\Xi^{\ast})}{\delta  \Xi(x)}{\rm exp
\mit}(-\frac{H(\Xi,\Xi^{\ast})}{T})\}}{\rm tr \mit (\rm exp
\mit(-\frac{H(\Xi,\Xi^{\ast})}{T}))}\nonumber\\
+\int \limits_0^1 ds \frac{{\{\rm tr \mit}K_m(\Xi,\Xi^{\ast}) {\rm
exp \mit(-s\frac{ H(\Xi,\Xi^{\ast})}{T}) }{\frac{\delta
H(\Xi,\Xi^{\ast})}{\delta\Xi(x)}}{\rm exp \mit(s\frac{
H(\Xi,\Xi^{\ast})}{T})}{\rm exp
\mit}(-\frac{H(\Xi,\Xi^{\ast})}{T})\}}{\rm tr \mit (\rm exp
\mit(-\frac{H(\Xi,\Xi^{\ast})}{T})) }\nonumber\\
-\frac{{\rm tr \mit}\{K_m(\Xi,\Xi^{\ast}){\rm exp
\mit}(-\frac{H(\Xi,\Xi^{\ast})}{T})\}}{\rm tr \mit (\rm exp
\mit(-\frac{H(\Xi,\Xi^{\ast})}{T})) }\frac{{\rm tr\{
\mit}{\frac{\delta H(\Xi,\Xi^{\ast})}{\delta\Xi(x)}}{\rm exp
\mit}(-\frac{H(\Xi,\Xi^{\ast})}{T})\}}{\rm tr \mit ({\frac{\delta
H(\Xi,\Xi^{\ast})}{\delta\Xi(x)}}\rm exp
\mit(-\frac{H(\Xi,\Xi^{\ast})}{T})) }
\end{eqnarray}
Let us show that each of three term in right hand side of this
equation is equal to zero. Let us consider the third term. It is
evident that this term is proportional to \(\dot{\Xi}(x,t)\). But
the soliton which has nontrivial intersection with \(C_i(\vec{a})\)
moves with zero velocity. So, this term is equal to zero. Let us
consider first term:
\begin{eqnarray}
\frac{{\rm tr \mit}\{\frac{\delta K_m(\Xi,\Xi^{\ast})}{\delta
\Xi(x)}{\rm exp \mit}(-\frac{H(\Xi,\Xi^{\ast})}{T})\}}{\rm tr \mit
(\rm exp \mit(-\frac{H(\Xi,\Xi^{\ast})}{T}))}.
\end{eqnarray}
This term is proportional to \(\langle\Xi'(x,t)\rangle\), where dash
means the derivative corresponding to the flow, which generates by
\(K_m(\Xi,\Xi^{\ast})\). The averaging here is an averaging over
generalized microcanonical distribution. The flow generated by
\(K_m(\Xi,\Xi^{\ast})\) preserve this microcanonical distribution.
Therefore \(\langle\Xi'(x,t)\rangle=0\).

At last let us consider second term. Let us represent it as follows:
\begin{eqnarray}
\sum \limits_{n=0}^{+\infty}\frac{1}{n!} \int \limits_0^1
(-\frac{s}{T})^n\frac{{\{\rm tr \mit}K_m(\Xi,\Xi^{\ast})
[H(\Xi,\Xi^{\ast}),...[H(\Xi,\Xi^{\ast}),{\frac{\delta
H(\Xi,\Xi^{\ast})}{\delta\Xi(x)}}]...]{\rm exp
\mit}(-\frac{H(\Xi,\Xi^{\ast})}{T})\}}{\rm tr \mit (\rm exp
\mit(-\frac{H(\Xi,\Xi^{\ast})}{T})) } \label{IterCommut}
\end{eqnarray}
Note that the iterated commutator here remains finite if the volume
of \(C_i(\vec{a})\) tends to infinity. Note that the averaging in
last formula is an averaging over the generalized microcanonical
distribution:
\begin{eqnarray}
\rho=\rm const \mit \delta(H-E) \prod \limits_{j=1}^N
\delta(K_j-K'_j)
\end{eqnarray}
In result in each term of series (\ref{IterCommut}) the integral
\(K_mm(\Xi,\Xi^{\ast})\) could be replaced by its observable value
\(K'_m\). In result the second term is equal to
\begin{eqnarray}
\frac{K'_m \rm tr \mit (\frac{\delta
H(\Xi,\Xi^{\ast})}{\delta\Xi(x)}\rm exp
\mit(-\frac{H(\Xi,\Xi^{\ast})}{T}))}{\rm tr \mit (\rm exp
\mit(-\frac{H(\Xi,\Xi^{\ast})}{T})) }
\end{eqnarray}
One can prove that this term is equal to zero by the same method as
has been used for third term. Therefore the theorem is proved.

\textbf{Programm --- Hypothesis.} So we have proved that the
equation for condensate wave function (hamiltonian with respect to
\(F(\Xi,\Xi^{\ast}|T)\)) admit a lot of independent first integrals
in involution (in limit \(\lambda=0,\;T=0\)). In the limit
\(\lambda=0,\;T=0\) the condensate wave function (after suitable
rescaling of space coordinates and time) satisfies to Pitaevskii
equation. Pitaevskii equation is a non linear Schrodinger equation.
In \(1D\) case the nonlinear Pitaevskii equation is an completely
integrable system. But we have proved that in \(3D\) case nonlinear
Schrodinger equation (under the assumption of asymptotical
completeness) admit a lot of independent integrals in involution.
Therefore there arise the following program-hypothesis of
constructing (not necessary completely) integrable systems. One can
take some system of statistical mechanics such that there exists a
phase transition of second order at some temperature concerning with
some symmetry breaking. Then one may find some small parameter
\(\eta\) for this system and write some (partial) differential
equation for the order parameter in the limit \(\lambda=0\) by some
exact asymptotical method. Obtained equation will be a hamiltonian
system which admit a lot of independent commuting first integrals in
involution.

\section{Conclusion.} In the present paper we have considered some
physical examples to illustrate basics principles of the generalized
thermodynamics, developed in [2]. I am very grateful to A.V. Koshelkin for
valuable critical comments on this article and very useful discussions.

\end{document}